\documentclass[prd,eqsecnum,amsfonts,twocolumn,groupedaddress,nofootinbib,twosided,showkeys,tighten]{revtex4}
\usepackage{graphicx}
\usepackage{amssymb}
\usepackage{bm}
\usepackage[colorlinks]{hyperref}

\usepackage{color}
\definecolor{IITred}{rgb}{0.5,0.05,0.05}

\newcommand{\onetev}{1-TeV scale}

\newcommand{\ie}{{\em i.e.}}
\newcommand{\cf}{{\em cf.\ }}

\newcommand{\gev}{\hbox{ GeV}}

\newcommand{\ev}{\hbox{ eV}}
\newcommand{\mev}{\hbox{ MeV}}

\newcommand{\tev}{\hbox{ TeV}}

\newcommand{\cm}{\hbox{ cm}}
\newcommand{\mm}{\hbox{ mm}}

\newcommand{\pb}{\hbox{ pb}}
\newcommand{\fb}{\hbox{ fb}}

\newcommand{\m}{\hbox{ m}}

\newcommand{\eqn}[1]{(\ref{#1})}
\newcommand{\Eqn}[1]{Eq.~(\ref{#1})}
\newcommand{\abs}[1]{\left| #1\right|}
\newcommand{\smgg}{\ensuremath{\mathrm{SU(3)_c}\otimes \mathrm{SU(2)_L}\otimes \mathrm{U(1)}_Y}}
\newcommand{\lrgg}{\ensuremath{\mathrm{SU(3)_c}\otimes \mathrm{SU(2)_L}\otimes \mathrm{SU(2)_R}\otimes \mathrm{U(1)}_{B-L}}}

\newcommand{\ewgg}{\ensuremath{\mathrm{SU(2)_L}\otimes \mathrm{U(1)}_Y}}

\newcommand{\wigg}{\ensuremath{\mathrm{SU(2)_L}}}
\newcommand{\emgg}{\ensuremath{\mathrm{U(1)}_{\mathrm{em}}}}

\def\vev#1{\left\langle #1\right\rangle_0}
\newcommand{\cfrac}[2]{\ensuremath{\textstyle{\frac{#1}{#2}}}}

\begin{document}
\preprint{FERMILAB--PUB--09/230--T}
\thispagestyle{empty}
\title{Unanswered Questions in the Electroweak Theory}

\markboth{Chris Quigg}{Unanswered Questions in the Electroweak Theory}

\author{Chris Quigg}\email{quigg@fnal.gov}
\affiliation{Theoretical Physics Department, Fermi National Accelerator Laboratory, 
 Batavia, Illinois 60510 USA  \\
Institut f\"{u}r Theoretische Teilchenphysik, Universit\"{a}t Karlsruhe, D-76128 Karlsruhe, Germany
\\ Theory Group, Physics Department, CERN, CH-1211 Geneva 23, Switzerland}

\begin{keywords}
 {Electroweak symmetry breaking, Higgs boson, 1-TeV scale, Large Hadron Collider, hierarchy problem, extensions to the standard model\hfill \fbox{\textsf{FERMILAB--PUB--09/230--T}}}
\end{keywords}

\begin{abstract}
This article is devoted to the status of the electroweak theory on the eve of experimentation at CERN's Large Hadron Collider. A compact summary of the logic and structure of the electroweak theory precedes an examination of  what experimental tests have established so far. The outstanding unconfirmed prediction of the electroweak theory is the existence of the Higgs boson, a weakly interacting spin-zero particle that is the agent of electroweak symmetry breaking, the giver of mass to the weak gauge bosons, the quarks, and the leptons. General arguments imply that the Higgs boson or other new physics is required on the TeV energy scale. Indirect constraints from global analyses of electroweak measurements suggest that the mass of the standard-model Higgs boson is less than $200\gev$. Once its mass is assumed, the properties of the Higgs boson follow from the electroweak theory, and these inform the search for the Higgs boson. Alternative mechanisms for electroweak symmetry breaking are reviewed, and the importance of electroweak symmetry breaking is illuminated by considering a world without a specific mechanism to hide the electroweak symmetry.

For all its triumphs, the electroweak theory has many shortcomings. It does not make specific predictions for the masses of the quarks and leptons or for the mixing among different flavors. It leaves unexplained how the Higgs-boson mass could remain below $1\tev$ in the face of quantum corrections that tend to lift it toward the Planck scale or a unification scale. The Higgs field that must pervade all of space to hide the electroweak symmetry contributes a vacuum energy density far in excess of what is observed. And the electroweak theory responds inadequately to challenges raised by astronomical observations, including the dark-matter problem and the baryon asymmetry of the Universe. These shortcomings argue for physics beyond the standard model; some possibilities are recalled.

The Large Hadron Collider moves experiments squarely into the TeV scale, where answers to important outstanding questions will be found. The questions are developed in the course of this review, and a short summary is attempted of how knowledge might accumulate.
\end{abstract}

\maketitle

\section{INTRODUCTION \label{sec:intro}}
The electroweak theory~\cite{Glashow:1961tr,Weinberg:1967tq,Salam} joins electromagnetism with the weak force in a single relativistic quantum field theory. Electromagnetism is a force of infinite range, while the influence of the charged-current weak interaction responsible for radioactive beta decay only spans distances shorter than about $10^{-15}\cm$, less than 1\% of the proton radius.  The two interactions, so different in their range and apparent strength, are ascribed to a common gauge symmetry. We say that the electroweak gauge symmetry is spontaneously broken to the  gauge symmetry of electromagnetism.
Through the past two decades, precision experiments have elevated the electroweak theory from a promising description to a provisional law of nature, tested as a quantum field theory at the level of one percent or better by many measurements.
Joined with quantum chromodynamics, the theory of the strong interactions, to form the \textit{standard model}, and augmented to incorporate neutrino masses and lepton mixing, it describes a vast array of experimental information.
The development and validation of the standard model is a landmark in the history of science.

One measure of the sweep of the electroweak theory is that its predictions hold over a prodigious range of distances, from about $10^{-18}\m$ to more than $10^8\m$. The origins of the theory lie in the discovery of Coulomb's law in tabletop experiments by Cavendish and Coulomb. It was stretched to longer and shorter distances by the progress of experiment. In the long-distance limit, the classical electrodynamics of a massless photon suffices. At shorter distances than the human scale, classical electrodynamics was superseded by quantum electrodynamics (QED), which is now subsumed in the electroweak theory, tested at energies up to a few hundred GeV. Some key steps in the evolution may be traced in~\cite{Hung:1980rh,Weinberg:2008zzb}.

The electroweak theory anticipated the existence and properties of weak neutral-current interactions, predicted the properties of the gauge bosons $W^\pm$ and $Z^0$ that mediate charged-current and neutral-current interactions, and required the fourth quark flavor, charm. Fits to a universe of electroweak precision measurements are in excellent agreement with the standard model. 

How the electroweak gauge symmetry is spontaneously broken is one of the most urgent and challenging questions before particle physics. The standard-model answer is an elementary scalar field whose self-interactions select a vacuum state in which the full electroweak symmetry is hidden. However, the Higgs boson, as the elementary scalar is known,  has not been observed directly, and we do not know whether a fundamental Higgs field exists or a different agent breaks electroweak symmetry. Finding the Higgs boson or its replacement is one of the great campaigns now under way in both experimental and theoretical particle physics.

The aim of this article is to survey what we know and what we need to know about the electroweak theory, in anticipation of the experiments soon to begin at the Large Hadron Collider, a high-luminosity proton-proton machine that will reach $14\tev$ c.m.\ energy. We begin in \S\ref{sec:ewtheory} with a short summary of the essential elements of the electroweak theory. Next, in \S\ref{sec:law}, we will examine the experimental support that has helped to establish the electroweak theory. The evidence includes the behavior of the couplings at the Lagrangian level, along with signs for weak-electromagnetic unification. A prominent feature of the electroweak theory is the absence of flavor-changing neutral currents. An important chapter in the weak interactions, just concluded, validated the picture of three-family quark mixing that organizes a vast amount of experimental information, including the observations of \textsf{CP} violation. Quantum corrections test the electroweak theory as a quantum field theory and give evidence for the interactions of (something resembling) the Higgs boson with the weak gauge bosons. Low-energy tests of the electroweak theory can be expressed as determinations of the weak mixing parameter. The electroweak theory gives but a partial explanation for the origin of quark and lepton masses, so I regard all the quark and lepton masses as evidence for physics beyond the standard model.

The Higgs boson, the missing ingredient of the standard model, is the subject of \S\ref{sec:agent}. There we describe theoretical and experimental constraints on the Higgs-boson mass and outline the production and decay characteristics that will govern the search at the LHC. Alternatives to the Higgs mechanism, beginning with dynamical symmetry breaking inspired by the microscopic theory of the superconducting phase transition, are described. I devote a brief passage to what the world would have been like, in the absence of an explicit mechanism to hide the electroweak symmetry. This excursion underlines the importance of discovering the agent of electroweak symmetry breaking for our understanding of the everyday world.

\S\ref{sec:SMincomplete} is devoted to the shortcomings of the standard model, including the partial understanding of fermion masses and mixing among quark families, the challenge of stabilizing the Higgs mass below $1\tev$ in the face of quantum corrections, and the vacuum energy problem. We take note of questions that lie beyond the scope of the standard model: the nature of dark matter, the matter asymmetry of the universe, the quantization of electric charge, and the role of gravity. Both sets of issues motivate more complete and predictive extensions to the standard model.

The new era ushered in by the Large Hadron Collider is the subject of \S\ref{sec:dawn}. I pose a series of electroweak questions for the LHC, and then note some possibilities for new physics motivated by the hierarchy problem and the search for dark-matter candidates. I describe how new knowledge might build up as the LHC data samples grow, and remark on the continuing role of experiments at the intensity frontier. A short summary concludes the article in \S\ref{sec:sum}.

\section{THE ELECTROWEAK THEORY \label{sec:ewtheory}}
The electroweak theory, and the path by which it evolved, is developed in many modern textbooks, including~\cite{Aitchison,CGExpFound,ChengLi,DonoghueGH,Pesky,CQFIP56}. Useful perspectives on the current situation are presented in lecture courses, including~\cite{Altarelli:2008fd,Dawson:2005dv,Langacker:2009my}. Here we give a quick summary of the essential ideas and outcomes.

 We build the standard model of particle physics on a set of constituents that we regard provisionally as elementary: the quarks and leptons,
plus a few fundamental forces derived from gauge symmetries. The quarks 
are influenced by the strong interaction, and so carry \textit{color}, 
the strong-interaction charge, whereas the leptons do not feel the 
strong interaction, and are colorless. We idealize the quarks and leptons as pointlike, because they  show no evidence of internal structure at 
the current limit of our resolution,  ($r \lesssim 10^{-18}\m$). The charged-current weak interaction responsible for radioactive beta decay and other processes acts only on the left-handed fermions. Whether the observed parity violation reflects a fundamental asymmetry in the laws of Nature, or a left-right symmetry that is hidden by circumstance and might be restored at higher energies, we do not know. 

Like its forerunner, quantum electrodynamics, the electroweak theory
is a gauge theory, in which interactions follow from symmetries. The correct electroweak gauge symmetry, which melds an $\mathrm{SU(2)_{L}}$ family (weak-isospin) symmetry with a $\mathrm{U(1)}_{Y}$ weak-hypercharge phase symmetry, emerged through trial and error, guided by experiment. We characterize the leptonic sector of the \ewgg\ theory by  the left-handed leptons
\begin{equation}
\mathsf{L}_e = 
\left(
		\begin{array}{c}
			\nu_{e}  \\
			e^{-}
		\end{array}
		 \right)_{\mathrm{L}} \;\;
\mathsf{L}_{\mu} = 
		\left(
		\begin{array}{c}
			\nu_{\mu}  \\
			\mu^{-}
		\end{array}
		 \right)_{\mathrm{L}} \;\;
\mathsf{L}_{\tau} = 
		\left(
		\begin{array}{c}
			\nu_{\tau}  \\
			\tau^{-}
		\end{array}
		\right)_{\mathrm{L}}	\;,
		\label{eq:lleptons}
	\end{equation}
with weak isospin $I = \cfrac{1}{2}$ and weak hypercharge $Y(\mathsf{L}_{\ell}) = -1$, and the right-handed weak-isoscalar charged leptons
\begin{equation}
\mathsf{R}_{e,\mu,\tau} = e_{\mathrm{R}}, \mu_{\mathrm{R}}, \tau_{\mathrm{R}}\;,
\label{eq:rightlep}
\end{equation}
with weak hypercharge $Y(\mathsf{R}_{\ell}) = -2$. The weak hypercharges are chosen to reproduce the observed electric charges, through the connection $Q = I_{3} + \cfrac{1}{2}Y$. Here we have idealized the neutrinos as massless. Very brief comments on massive neutrinos will be found in \S\ref{subsec:ID}.

The hadronic sector consists of the left-handed quarks
\begin{equation}
\mathsf{L}_q^{(1)} = 
\left(
		\begin{array}{c}
			u  \\
			d^\prime
		\end{array}
		 \right)_{\mathrm{L}} \;\;
\mathsf{L}_q^{(2)} = 
		\left(
		\begin{array}{c}
			c  \\
			s^\prime
		\end{array}
		 \right)_{\mathrm{L}} \;\;
		 \mathsf{L}_q^{(3)} = 
		\left(
		\begin{array}{c}
			t  \\
			b^\prime
		\end{array}
		 \right)_{\mathrm{L}}	\;,
		 \label{eq:lquarks}
	\end{equation}	
with weak isospin $I = \cfrac{1}{2}$ and weak hypercharge $Y(\mathsf{L}_q) = \cfrac{1}{3}$, and their right-handed weak-isoscalar counterparts
\begin{equation}
\mathsf{R}_u^{(1,2,3)} = u_{\mathrm{R}}, c_{\mathrm{R}}, t_{\mathrm{R}}\hbox{ and }
\mathsf{R}_d^{(1,2,3)} = d_{\mathrm{R}}, s_{\mathrm{R}}, b_{\mathrm{R}}\;,
\label{eq:rightup}
\end{equation}
with weak hypercharges $Y(\mathsf{R}_u) = \cfrac{4}{3}$ and $Y(\mathsf{R}_d) = -\cfrac{2}{3}$.
The primes on the lower components of the quark doublets in (\ref{eq:lquarks}) signal that the weak eigenstates are mixtures of the mass eigenstates:
\begin{equation}
\left(\begin{array}{c} d^\prime \\ s^\prime \\ b^\prime \end{array}\right) = \left( \begin{array}{ccc} 
V_{ud} & V_{us} & V_{ub} \\
V_{cd} & V_{cs} & V_{cb} \\
V_{td} & V_{ts} & V_{tb} 
\end{array}\right) \left(\begin{array}{c} d \\ s \\ b \end{array}\right)  \equiv \mathsf{V} \left(\begin{array}{c} d \\ s \\ b \end{array}\right) ,
\label{eq:ckmmatrix}
\end{equation}
where the $3 \times 3$ unitary Cabibbo~\cite{Cabibbo:1963yz}--Kobayashi--Maskawa~\cite{Kobayashi:1973fv} matrix $\mathsf{V}$ expresses the quark mixing. See \S\ref{subsec:ckmtests} for further discussion.

The fact that each left-handed lepton doublet is matched by a left-handed quark doublet guarantees that the theory is anomaly free, so that quantum corrections respect the gauge symmetry~\cite{Bouchiat:1972iq}.

The \ewgg\ electroweak gauge group entails two sets of gauge fields: a weak isovector $\bm{b}_\mu$, with coupling constant $g$, and a weak isoscalar ${{\mathcal A}}_\mu$, with its own coupling constant $g^\prime$. The gauge fields compensate for the variations induced by gauge transformations, provided that they obey the transformation laws $\bm{b}_\mu \to \bm{b}_\mu - \bm{\alpha} \times \bm{b}_\mu - (1/g)\partial_\mu \bm{\alpha}$ under an infinitesimal weak-isospin rotation generated by $G = 1 + (i/2)\bm{\alpha} \cdot \bm{\tau}$ (where $\bm{\tau}$ are the Pauli isospin matrices) and $\mathcal{A}_\mu \to \mathcal{A}_\mu - (1/g^\prime)\partial_\mu \alpha$ under an infinitesimal hypercharge phase rotation.
Corresponding to these gauge fields are the field-strength tensors 
\begin{equation}
    F^{\ell}_{\mu\nu} = \partial_{\nu}b^{\ell}_{\mu} - 
    \partial_{\mu}b^{\ell}_{\nu} + 
    g\varepsilon_{jk\ell}b^{j}_{\mu}b^{k}_{\nu}\; ,
    \label{eq:Fmunu}
\end{equation}
($\ell = 1,2,3$) for the weak-isospin symmetry, and 
\begin{equation}
    f_{\mu\nu} = \partial_{\nu}{{\mathcal A}}_\mu - \partial_{\mu}{{\mathcal 
    A}}_\nu \; , 
    \label{eq:fmunu}
\end{equation}
for the weak-hypercharge symmetry. 
 
We may summarize the interactions 
by the Lagrangian
\begin{equation}
\mathcal{L} = \mathcal{L}_{\rm gauge} + \mathcal{L}_{\rm leptons} +  \mathcal{L}_{\rm quarks}\ ,                           
\end{equation}             
with
\begin{equation}
\mathcal{L}_{\rm gauge}=-\cfrac{1}{4}\sum_\ell {F}^\ell_{\mu\nu}  {F}^{\ell\,\mu\nu}
-\cfrac{1}{4}f_{\mu\nu}f^{\mu\nu},
\label{eq:gaugeL}
\end{equation}
\begin{eqnarray}     
\mathcal{L}_{\rm leptons} & = & \overline{{\sf R}}_{\ell}\:i\gamma^\mu\!\left(\partial_\mu
+i\frac{g^\prime}{2}{\cal A}_\mu Y\right)\!{\sf R}_{\ell}
\label{eq:matiere} \\ 
& + & \overline{{\sf
L}}_{\ell}\:i\gamma^\mu\!\left(\partial_\mu 
+i\frac{g^\prime}{2}{\cal
A}_\mu Y+i\frac{g}{2}\bm{\tau}\cdot\bm{b}_\mu\right)\!{\sf L}_{\ell}\;, \nonumber
\end{eqnarray}
where $\ell$ runs over $e, \mu, \tau$, and
\begin{eqnarray}     
\mathcal{L}_{\rm quarks} & = & \overline{{\sf R}}_{u}^{(n)}\:i\gamma^\mu\!\left(\partial_\mu
+i\frac{g^\prime}{2}{\cal A}_\mu Y\right)\!{\sf R}_{u}^{(n)}
\nonumber \\ 
 & + & \overline{{\sf R}}_{d}^{(n)}\:i\gamma^\mu\!\left(\partial_\mu
+i\frac{g^\prime}{2}{\cal A}_\mu Y\right)\!{\sf R}_{d}^{(n)}
\label{eq:qmatiere}  \\
& + & \overline{{\sf
L}}_{q}^{(n)}\:i\gamma^\mu\!\left(\partial_\mu 
+i\frac{g^\prime}{2}{\cal
A}_\mu Y+i\frac{g}{2}\bm{\tau}\cdot\bm{b}_\mu\right)\!{\sf L}_{q}^{(n)}\;, \nonumber
\end{eqnarray}
where the generation index $n$ runs over $1, 2, 3$. The objects in parentheses in \eqn{eq:matiere} and \eqn{eq:qmatiere} are the \textit{gauge-covariant derivatives.}

Although the weak and electromagnetic interactions share a common origin in the \ewgg\ gauge symmetry, their manifestations are very different. Electromagnetism is a force of infinite range, while the influence of the charged-current weak interaction responsible for radioactive beta decay only spans distances shorter than about $10^{-15}\cm$. The established phenomenology of the weak interactions is thus at odds with the theory we have developed to this point. The gauge Lagrangian (\ref{eq:gaugeL}) contains four massless electroweak gauge bosons, \textit{viz.\/} $b^{1}_{\mu}$, $b^{2}_{\mu}$, $b^{3}_{\mu}$, ${{\mathcal A}}_\mu$. They are massless because a mass term such as $\cfrac{1}{2}m^2\mathcal{A}_\mu\mathcal{A}^\mu$ is not invariant under a gauge transformation. Nature has but one: the photon. Moreover, the \ewgg\ gauge symmetry forbids fermion mass terms $m\bar{f}\!f = m(\bar{f}_{\mathrm{R}}f_{\mathrm{L}} + \bar{f}_{\mathrm{L}}f_{\mathrm{R}})$ in (\ref{eq:matiere}) and (\ref{eq:qmatiere}), because the left-handed and right-handed fields transform differently. 

To give masses to the gauge bosons and constituent fermions, we must hide the electroweak symmetry, recognizing that a symmetry of the laws of Nature does not imply that the same symmetry will be manifest in the outcomes of those laws. 
   
The superconducting phase transition offers an instructive model for hiding the electroweak gauge symmetry. To give masses to the intermediate bosons of the weak interaction, we appeal to the Meissner effect---the exclusion of magnetic fields from a superconductor, which corresponds to the photon developing a nonzero mass within the superconducting medium. What has come to be called the Higgs mechanism~\cite{Englert:1964et,Higgs:1964ia,Higgs:1964pj,Guralnik:1964eu} can be understood as a relativistic generalization of the Ginzburg-Landau phenomenology~\cite{Ginzburg:1950sr} of superconductivity. 

Let us see how spontaneous symmetry breaking operates in the electroweak theory.
We introduce a complex doublet of scalar fields
\begin{equation}
\phi\equiv \left(\begin{array}{c} \phi^+ \\ \phi^0 \end{array}\right)
\end{equation}
with weak hypercharge $Y_\phi=+1$.  Next, we add to the Lagrangian new 
(gauge-invariant) terms for the interaction and propagation of the 
scalars,
\begin{equation}
      \mathcal{L}_{\rm scalar} = (\mathcal{D}^\mu\phi)^\dagger(\mathcal{D}_\mu\phi) - V(\phi^\dagger \phi),
      \label{eq:scalarpot}
\end{equation}
where the gauge-covariant derivative is
\begin{equation}
      \mathcal{D}_\mu=\partial_\mu 
+i\frac{g^\prime}{2}{\cal A}_\mu
Y+i\frac{g}{2}\bm{\tau}\cdot\bm{b}_\mu \; ,
\label{eq:GcD}
\end{equation}
and (inspired by Ginzburg \& Landau) the potential interaction has the form
\begin{equation}
      V(\phi^\dagger \phi) = \mu^2(\phi^\dagger \phi) +
\abs{\lambda}(\phi^\dagger \phi)^2 .
\label{eq:SSBpot}
\end{equation}
We are also free to add  gauge-invariant Yukawa interactions between the scalar fields
and the leptons ($\ell$ runs over $e, \mu, \tau$ as before),
\begin{equation}
      \mathcal{L}_{{\rm Yukawa-}\ell} = -\zeta_{\ell}\left[(\overline{{\sf L}}_{\ell}\phi){\sf R}_{\ell} + 
      \overline{{\sf R}}_{\ell}(\phi^\dagger{\sf
L}_{\ell})\right]\;,
\label{eq:Yukterm}
\end{equation}
and similar interactions with the quarks.

Now we arrange 
the self-interactions of the scalars so that the vacuum state corresponds to a 
broken-symmetry solution.  The electroweak symmetry is spontaneously broken if the parameter
$\mu^2$ is taken to be negative. In that event, gauge invariance gives us the freedom to choose the state of minimum energy---the vacuum state---to correspond to the vacuum expectation value
\begin{equation}
\vev{\phi} = \left(\begin{array}{c} 0 \\ v/\sqrt{2} \end{array}
\right),
\label{eq:vevis}
\end{equation}
where $v = \sqrt{-\mu^2/\abs{\lambda}}$.

The vacuum of \eqn{eq:vevis} breaks the gauge symmetry $\ewgg \to \mathrm{U(1)}_{\mathrm{em}}$.  The vacuum state $\vev{\phi}$ is invariant under a symmetry operation corresponding to the generator ${\mathcal G}$ provided that $e^{i \alpha {\mathcal G}}\vev{\phi} = \vev{\phi}$, \ie, if ${\mathcal G}\vev{\phi} = 0$.  
Direct calculation reveals that  the  original four generators are all broken, but electric charge is
not.  The photon remains massless, but the other three gauge bosons acquire 
masses, as auxiliary scalars assume the role of the third 
(longitudinal) degrees of freedom.  

Introducing the weak mixing angle $\theta_{\mathrm{W}}$ through the definition $g^{\prime} = g\tan\theta_{\mathrm{W}}$, we can express the photon as the linear combination
$A = \mathcal{A}\cos{\theta_{\mathrm{W}}} + b_{3}\sin{\theta_{\mathrm{W}}}$. We identify the strength of its (pure vector) coupling to charged particles, $gg^{\prime}/\sqrt{g^{2} + g^{\prime 2}}$, with the electric charge $e$.
The mediator of the charged-current weak 
interaction, $W^{\pm} = (b_{1} \mp ib_{2})/\sqrt{2}$, acquires a 
mass 
\begin{equation}
M_{W} = gv/2 = ev/2\sin{\theta_{\mathrm{W}}} .
\label{eq:wmass}
\end{equation}
The electroweak gauge theory reproduces the low-energy phenomenology of the $V - A$ theory of weak interactions, provided we set $v = (G_{\mathrm{F}}\sqrt{2})^{-1/2} = 246\gev$, where $G_{\mathrm{F}} = 1.166 37(1) \times 10^{-5}\gev^{-2}$ is Fermi's weak-interaction coupling constant. It follows at once that $M_W \approx 37.3\gev/\sin{\theta_{\mathrm{W}}}$. The combination of the $I_3$ and $Y$ gauge bosons orthogonal to the photon is the mediator of the neutral-current weak interaction, $Z = b_{3}\cos{\theta_{\mathrm{W}}} - \mathcal{A}\sin{\theta_{\mathrm{W}}}$, which
acquires a mass 
\begin{equation}
M_Z=M_W/\cos{\theta_{\mathrm{W}}} .
\label{eq:zmass}
\end{equation} 

The masses of the elementary fermions are not predicted by the electroweak theory. Each fermion mass involves a new Yukawa coupling $\zeta$ (\cf \eqn{eq:Yukterm}).  
When the electroweak symmetry is spontaneously broken, the electron 
mass emerges as $m_{e} = \zeta_{e}v/\sqrt{2}$. The Yukawa couplings that reproduce the observed quark and lepton masses range over many orders of magnitude, as detailed in \S\ref{subsec:fmassgen}. We do not know what sets the values of the Yukawa couplings. They do not follow from a known symmetry principle, for example.

Three of the four scalar degrees of freedom that we introduced to contrive a vacuum state that hides the electroweak gauge symmetry have become the longitudinal components of $W^+$, $W^-$, and $Z$. The fourth appears as a massive
spin-zero particle, called the Higgs boson, $H$, a vestige 
of the spontaneous symmetry breaking.  Its mass  is 
given symbolically as $M_{H}^{2} = -2\mu^{2} > 0$, but \textit{we have no 
prediction for its value.}  On the other hand, the interactions of the Higgs boson with gauge bosons and fermions are completely specified---after spontaneous symmetry breaking---by the Lagrangian terms $\mathcal{L}_{\mathrm{scalar}}$ and $\mathcal{L}_{\mathrm{Yukawa}}$. Given the mass of the Higgs boson, we may calculate its properties.

Let us summarize how particle mass arises in the standard electroweak theory. Unless the electroweak gauge symmetry is hidden, the four gauge bosons and all the constituent fermions are massless. Spontaneous symmetry breaking, in the form of the Higgs mechanism, gives masses to the weak gauge bosons and creates the possibility for the fermions to acquire mass. Once the weak mixing parameter $\sin^2\theta_{\mathrm{W}}$ is fixed by the study of weak-neutral-current interactions, the theory makes successful quantitative predictions for the $W^\pm$- and $Z$-boson masses. Although the natural scale of fermion masses would seem to be set by the electroweak scale, the specific values are determined by Yukawa couplings of the fermions to the Higgs field. These Yukawa couplings are not predicted by the electroweak theory. Finally, the theory requires a scalar Higgs boson, but does not make an explicit prediction for its mass.

\section{HOW THE ELECTROWEAK THEORY BECAME A LAW OF NATURE  \ldots 
 AND WHAT WE REALLY KNOW \label{sec:law}}
The \ewgg\ electroweak theory was formulated in the context of extensive experimental information about the charged-current weak interactions. Central elements included the parity-violating $V-A$ structure of the charged current and the Cabibbo universality of leptonic and semileptonic processes. On the theoretical front, a classic unitarity argument~\cite{Lee:1965js} made it clear that Fermi's four-fermion description could not be valid above c.m.\ energy $\sqrt{s}=620\gev$. Analysis of the reaction $\nu\bar{\nu} \to W^+ W^-$ showed that the \textit{ad-hoc} introduction of intermediate vector bosons, to make the weak interaction nonlocal, had divergence diseases of its own~\cite{GellMann:1969wt}.

The weak neutral-current interaction was not detected before the electroweak theory was formulated. The prediction of this new phenomenon and the availability of high-energy neutrino beams spurred the search for experimental manifestations of the weak neutral current.  Its discovery in 1973~\cite{Hasert:1973ff,Haidt:2004ne} marked an important milestone, as did the observation a decade later~\cite{Rubbia:1985pv} of the $W^{\pm}$~\cite{Arnison:1983rp,Banner:1983jy} and $Z^0$~\cite{Arnison:1983mk,Bagnaia:1983zx} bosons. The early years were marked by some inconsistent experimental results and the invention of many alternatives to the \ewgg\ theme. How physicists sorted out the correct electroweak theory is a fascinating story, but off our topic here. We shall concentrate instead on the evidence that now tests and validates the electroweak theory. See~\cite{Langacker:2001jc} for a compact authoritative rendering of the role of precision measurements in establishing the electroweak theory as a law of nature.
 
 \subsection{Tree-level}
 Following the discovery of neutral-current interactions, the new phenomenon was taken up in a number of $\nu N$ and $\nu e$ scattering experiments. Despite their statistical limitations, the neutrino-electron scattering experiments helped guide the convergence to the \ewgg\ standard model. Under modest universality assumptions, the $\nu e$ cross section measurements, combined with measurements of the forward-backward asymmetry that arises from $\gamma\hbox{-}Z$ interference in the reaction $e^+e^- \to \mu^+\mu^-$, uniquely selected the \ewgg\ chiral couplings of $Z$ to charged leptons~\cite{Panman:1984kt}. Only a short time later, it was reasonable to proclaim that the chiral couplings to all the known quarks and leptons had been uniquely determined, in agreement with the \ewgg\ theory~\cite{Langacker:1988fp}.
 
 Along the way, delicate observations of parity-violating phenomena in atomic physics began to add complementary information. Studies of polarized electron-deuteron scattering~\cite{Prescott:1978tm} confirmed that the neutral-current interactions are parity violating, also ruling in favor of the standard model.
   
This impressive progress, punctuated by the discoveries of $W$ and $Z$, was prelude to the incisive experiments at the SLAC and CERN $Z$ factories.  Measurements of the $Z$ lineshape and a determination of the ``invisible'' width of the $Z$ confirmed the hypothesis that three generations of light neutrinos are present in neutral-current interactions. The current inference from the invisible width, of $2.985 \pm 0.009$ active light neutrino species~\cite{EWReview} is not only consistent with the three observed neutrino species, it leaves little room for decays of $Z$ into exotic weakly interacting particles.

The conclusion that only three active light species exist does not rule out a fourth generation of quarks and leptons, provided that the neutral leptons are heavy enough that their contributions to the invisible width would be negligible---if not zero! A fourth generation is constrained, but not excluded, by what we know of charged-current and neutral-current interactions~\cite{Holdom:2009rf}.

Many extensions to the electroweak theory predict the existence of one or more electrically neutral color-singlet $Z^\prime$ gauge bosons~\cite{Bogdan}. The most telling direct searches have been carried out at the Tevatron in searches for direct-channel ($q\bar{q} \to Z^\prime$) resonances in reactions such as $\bar{p}p \to \ell^+\ell^- + \hbox{anything}$. Translating experimental sensitivity into limits on the mass of a new neutral gauge boson is complicated by the fact that $Z^\prime$ couplings to fermions are model-dependent---in some cases, even generation dependent. For a representative collection of examples, the Tevatron searches imply that $M_{Z^\prime} \gtrsim 789\gev$ at 95\% CL. For a heavy clone of the standard-model $Z$ (its only virtue as an example is that it is easy to state), the 95\% CL bound is $M_{Z_{\mathrm{SM}}^\prime} > 1\,030\gev$~\cite{Aaltonen:2008ah}. Other searches look for evidence of  $Z^\prime \to W^+W^-$. Global fits to electroweak parameters and neutral-current studies away from the $Z$ pole are sensitive to a $Z^\prime$.
 
\begin{figure}[tb]
  \centerline{\includegraphics[width=\columnwidth]{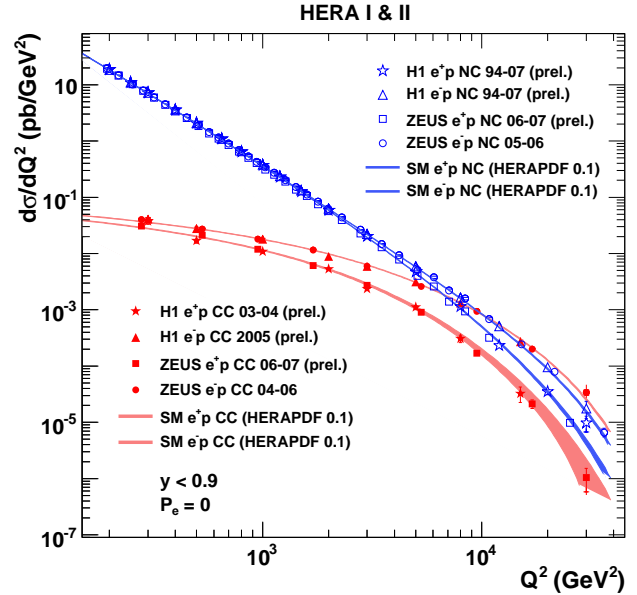}}
  \caption{The $Q^2$-dependence of the neutral-current (NC) and charged-current (CC) cross sections measured by the H1~\cite{H1prelim} and ZEUS~\cite{Chekanov:2008aa,Chekanov:2009gm,ZEUS09001,ZEUS09002} experiments at the HERA $e^\pm p$ collider. The curves represent the standard-model expectations derived from the HERA parton distribution functions. \label{fig:ewuni}}
 \end{figure}

The H1~\cite{H1prelim} and ZEUS~\cite{Chekanov:2008aa,Chekanov:2009gm,ZEUS09001,ZEUS09002} experiments at the $e^\pm p$ collider HERA compared the momentum-transfer dependence of neutral-current ($e^\pm p \to e^\pm + \hbox{anything}$) and charged-current ($e^\pm p \to (\bar{\nu}_e,\nu_e)  + \hbox{anything}$) at c.m.\ energies of $820\hbox{ and }920\gev$. A recent summary compiled by H1 and ZEUS is given in Figure~\ref{fig:ewuni}. At low-values of $Q^2$, the neutral-current cross section exceeds the charged-current cross section by more than two orders of magnitude, because the electromagnetic interaction is much stronger than the weak interaction at long wavelengths. For $Q^2 \gtrsim (M_W^2, M_Z^2)$, the cross sections roughly track each other. This behavior supports the notion that the intrinsic strengths of the weak and electromagnetic interactions are comparable.
 
 The absence of right-handed charged-current interactions is one of the foundational observations on which the \ewgg\ theory is built, and also a question that has lingered for more than fifty years. Is there a fundamental left-right asymmetry in the laws of nature, or did spontaneous symmetry breaking at some high scale give a large mass to a right-handed gauge boson, creating a low-energy preference for left-handed currents? This second possibility is the vision of left-right symmetric models, based on \lrgg\ gauge symmetry~\cite{Mohapatra:1974hk,Mohapatra:1974gc,Senjanovic:1975rk}. Searches for right-handed interactions, or for additional $W^{\prime\pm}$ gauge bosons are important probes of the electroweak theory~\cite{MuChun}.
 
 The direct searches at the Tevatron for $W^\prime \to e\nu$, assuming standard-model couplings, give a lower bound $M_{W^\prime} > 1\,000\gev$ at 95\% CL~\cite{Amsler:2008zzb}. A fit to low-energy data bounds the mass of a right-handed $W_{\mathrm{R}}$ as $M_{W_{\mathrm{R}}} > 715\gev$ at 90\%~CL, assuming that its gauge coupling is the same as the \wigg\ coupling, $g_{\mathrm{R}}=g_{\mathrm{L}}$~\cite{Czakon:1999ga}. Sensitive tests of the standard model are ongoing in  $\mu$ decay~\cite{Gagliardi:2005fg} and in $\beta$-decay~\cite{Severijns:2006dr}.

\begin{figure}[tb]
	\centerline{\includegraphics[width=\columnwidth]{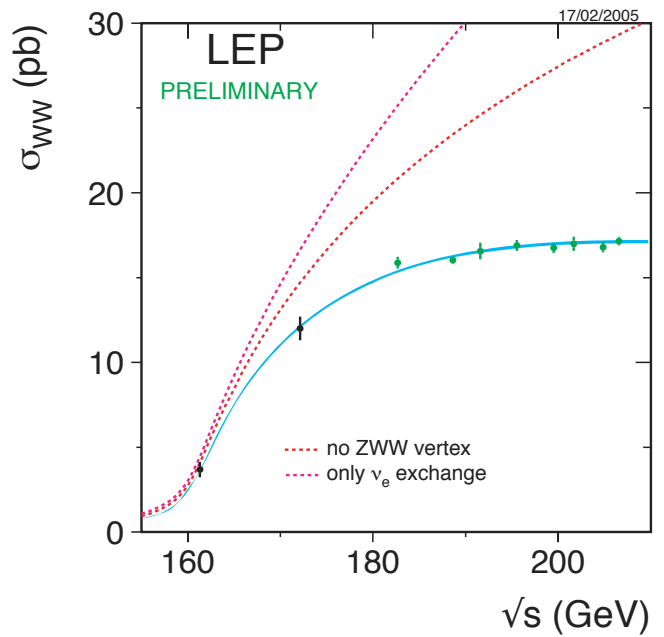}}
	\vspace*{6pt}
	\caption{Cross section for the reaction $e^{+}e^{-} \to W^{+}W^{-}$ 
	measured by the four LEP experiments, together with the full 
	electroweak-theory simulation and cross sections that would 
	result from $\nu$-exchange alone and from $(\nu+\gamma)$-exchange
	{\protect \cite{EWWG}}.}
	\protect\label{fig:LEPgc}
\end{figure}

 A noteworthy achievement of the LEP experiments is the validation of the $\mathrm{SU(2)_L \otimes U(1)}_Y$ symmetry for the interaction of gauge bosons with fermions and gauge bosons with gauge bosons in $e^+ e^- \to W^+ W^-$. This reaction is described by three 
Feynman diagrams that correspond to  $s$-channel photon and $Z^{0}$ exchange, and $t$-channel neutrino exchange. 
For the production of longitudinally polarized $W$-bosons, each diagram leads to a $J = 1$ partial-wave amplitude that grows as the square of the c.m.\ energy, but the gauge symmetry enforces a pattern of cooperation.
 The contributions of the direct-channel 
$\gamma$- and $Z^0$-exchange diagrams 
cancel the leading divergence in the $J=1$ 
partial-wave amplitude of the neutrino-exchange diagram.  The interplay is shown in Figure~\ref{fig:LEPgc}. If the $Z$-exchange contribution is omitted (middle line) or if both the 
$\gamma$- and $Z$-exchange contributions are omitted (upper line), the calculated cross section grows unacceptably with 
energy. The measurements compiled by the LEP Electroweak Working Group~\cite{EWWG} 
agree well with the benign high-energy behavior predicted by the full electroweak theory.

Tevatron measurements do not directly determine the $W^+W^-$invariant mass, because of the missing energy carried by neutrinos, but reach beyond the highest energy studied at LEP. The latest contributions, from the D0~\cite{Abazov:2009ys} and CDF~\cite{CDFWpair} Collaborations, are in agreement with standard-model expectations~\cite{Campbell:1999ah,Frixione:2002ik}, and tighten the bounds on anomalous couplings.

\subsection{Flavor-changing neutral currents \label{subsec:fcnc}}
Strangeness-changing neutral currents were the object of experimental searches even before the electroweak theory was conceived. It was recognized early on that  flavor-changing neutral-current effects cannot be isolated in nonleptonic decays. As an example, the transition $s \to d (u\bar{u})$ would be entangled with the charged-current transition $s \to u (d\bar{u})$. Accordingly, decays of hadrons into pairs of leptons have been the favored hunting ground for evidence of flavor-changing neutral currents (FCNC).

The branching fraction $\mathcal{B}(K^0_L \to \mu^+\mu^-) = (6.84 \pm 0.11)\times 10^{-9}$~\cite{Amsler:2008zzb} closely matches the standard expectation for decay through the (real and virtual) $\gamma\gamma$ intermediate state. The absence of strangeness-changing neutral-current interactions motivated Glashow, Iliopoulos, and Maiani~\cite{Glashow:1970gm} to advocate adding the charm quark $c$ to the then-familiar $u, d, s$, so that quark doublets
\begin{equation}
\left(\begin{array}{c} u \\ d\cos\theta_{\mathrm{C}} + s\sin\theta_{\mathrm{C}} \end{array} \right)_{\mathrm{L}}  \left(\begin{array}{c} c \\ s\cos\theta_{\mathrm{C}} - d\sin\theta_{\mathrm{C}} \end{array} \right)_{\mathrm{L}},
\label{eqn:GIMdoublets}
\end{equation}
where $\theta_{\mathrm{C}}$ is the Cabibbo angle, would mirror the then-known lepton doublets,
\begin{equation}
\left( \begin{array}{c} \nu_e \\ e \end{array}\right)_{\mathrm{L}}
\left( \begin{array}{c} \nu_\mu \\ \mu \end{array}\right)_{\mathrm{L}}\;.
\label{eqn:GIMleptons}
\end{equation}

The three-family generalization of the GIM mechanism banishes FCNC at lowest order, and greatly suppresses them at loop level~\cite{Buchalla:1995vs}. Verifying the absence of FCNC therefore tests the structure---and the completeness---of the electroweak theory. The most sensitive experimental search has been carried out in the $K^+ \to \pi^+ \nu \bar{\nu}$ channel.  Brookhaven Experiment 949 has observed three candidates, leading to a branching fraction $\mathcal{B}(K^+ \to \pi^+ \nu \bar{\nu})  = 1.73_{-1.05}^{+1.15} \times 10^{-10}$~\cite{Artamonov:2008qb}. This rate is consistent, within uncertainties, with the standard-model expectation, $\mathcal{B}(K^+ \to \pi^+ \nu \bar{\nu}) = (0.85 \pm 0.07) \times 10^{-10}$~\cite{Brod:2008ss}.

The limits on FCNC involving heavier flavors are less stringent, but nevertheless raise the question: if new physics is to reveal itself on the 1-TeV scale, why have we seen no sign of FCNC?

Within the standard model, the rate anticipated for the decay $D^0 \to \mu^+ \mu^-$ is very small: $\mathcal{B}(D^0 \to \mu^+ \mu^-) \gtrsim 4 \times 10^{-13}$~\cite{Burdman:2001tf}. The CDF Collaboration bounds $\mathcal{B}(D^0 \to \mu^+ \mu^-) < 5.3 \times 10^{-7}$ at 95\% C.L.~\cite{CDF9226}. For a general review of charmed meson decays, see ~\cite{Artuso:2008vf}. The observation of $D^0\hbox{-}\bar{D}^0$ mixing~\cite{Aubert:2007wf,Staric:2007dt} has intensified interest in the search for new physics in charmed-meson decays. Theoretical expectations are catalogued in~\cite{Burdman:2001tf,Burdman:2003rs,Bianco:2003vb,Golowich:2007ka,Golowich:2009ii,Bigi:2009df}.

An informative introduction to FCNC phenomena in $B$-meson decays is given in the \textit{BaBar Physics Book}~\cite{Harrison:1998yr}. The current experimental limit on leptonic $B_s$ decays,  $\mathcal{B}(B_s\to \mu^+ \mu^-) < 5.8 \times 10^{-8}$ at 95\% C.L.~\cite{Aaltonen:2007kv}, approaches standard-model sensitivity, $\mathcal{B}(B_s\to \mu^+ \mu^-) = (3.6 \pm 0.3) \times 10^{-9}$~\cite{Buras:2009us}. The corresponding limit for $B^0$ is $\mathcal{B}(B_d \to \mu^+ \mu^-) < 1.8 \times 10^{-8}$~\cite{Aaltonen:2007kv}, to be compared with the standard-model expectation, $\mathcal{B}(B_d \to \mu^+ \mu^-) = (1.1 \pm 0.1)\times 10^{-10}$. 


The world sample of top decays remains modest, and consequently the study of rare top decays is less advanced than for $K$, $D$, and $B$ mesons. From a search for single-top production, the CDF Collaboration reports $\mathcal{B}(t \to u g) < 3.9 \times 10^{-4}$ and $\mathcal{B}(t \to c g) < 5.7 \times 10^{-3}$ at 95\% C.L.~\cite{Aaltonen:2008qr}, improving earlier limits from LEP. The latter is to be compared with the standard-model expectation, $\mathcal{B}(t \to c g) \approx 10^{-10}$~\cite{Eilam:1990zc}. A study of top pair production yields $\mathcal{B}(t \to Zc) < 3.7\%$ at 95\% C.L.~\cite{Aaltonen:2008aaa}.

In charm and top decays, plenty of room remains  to search for physics beyond the standard model, as experiments approach standard-model sensitivity. But the absence of FCNC at tree level is firmly established. What we already know about (the suppression of) flavor-changing neutral current phenomena both challenges, and provides opportunities to uncover, many varieties of physics beyond the standard model, including dynamical electroweak symmetry breaking~\cite{Lane:2002wv,Hill:2002ap,Foadi:2007ue} and supersymmetry without auxiliary conditions~\cite{Masiero:1997bv}. The existing constraints have stimulated conjectures about ``minimal flavor violation''~\cite{D'Ambrosio:2002ex} and approximate generational symmetries~\cite{Appelquist:2003hn}. 

The search for FCNC effects in heavy quark decays is an example of how high-sensitivity studies at low energies can complement direct discovery physics at the LHC. Experimental searches for lepton-flavor violation offer another window on new physics in the neutral-current sector~\cite{Kuno:1999jp,deGouvea2009303}.

\subsection{Tests of the CKM Paradigm \label{subsec:ckmtests}}
A generation ago, the Cabibbo hypothesis~\cite{Cabibbo:1963yz} brought clarity to a wealth of information about semileptonic decays of mesons and hyperons. Transcribed to modern language, the charged-current interactions among the light quarks are specified by
\begin{equation}
\mathcal{L}_{\mathrm{CC}}^{(q)} = -\frac{g}{\sqrt{2}}\bar{u}_{\mathrm{L}}\gamma^\mu d_{\theta\mathrm{L}}W^+_ \mu + \mathrm{h.c.},
\label{eq:Cabcur}
\end{equation}
where $g$ is the \wigg\ gauge coupling and 
\begin{equation}
d_\theta = d\cos\theta_{\mathrm{C}} + s\sin\theta_{\mathrm{C}}.
\label{eq:Cabdown}
\end{equation}
The form \eqn{eq:Cabcur} matches the charged-current interaction among leptons,
\begin{equation}
\mathcal{L}_{\mathrm{CC}}^{(\ell)} = -\frac{g}{\sqrt{2}}\bar{e}_{\mathrm{L}}\gamma^\mu \nu_{\mathrm{L}}W^-_ \mu + \mathrm{h.c.},
\label{eq:oldlepcur}
\end{equation}
and so expresses the universality of the charged-current weak interactions. Tests of the Cabibbo universality hypothesis relating the strengths of  $u \leftrightarrow d$, $u \leftrightarrow s$, and $\nu \leftrightarrow e$ transitions are reviewed in~\cite{Cabibbo:2003cu}.

In a prescient paper, following the Glashow-Iliopoulos-Maiani~\cite{Glashow:1970gm} call for a fourth quark that would be the charged-current partner of the orthogonal combination $s_\theta = s\cos\theta_{\mathrm{C}} - d\sin\theta_{\mathrm{C}}$ but before the discovery of charm, Kobayashi \& Maskawa~\cite{Kobayashi:1973fv} generalized Cabibbo's hypothesis to three quark generations, in order to accommodate $\mathsf{CP}$ violation. Quark mixing is expressed by the $3 \times 3$ unitary matrix defined in \eqn{eq:ckmmatrix}, colloquially called the CKM matrix. Their key insight is that an $n \times n$ unitary matrix can be parametrized in terms of $n(n-1)/2$ real mixing angles and $(n-1)(n-2)/2$ complex phases, after the freedom to redefine the phases of quark fields has been taken into account. The phase angle present in the $3 \times 3$ case could, they suggested, account for $\mathsf{CP}$ violation.

This simple conjecture has far-reaching implications~\cite{BigiSanda,BrancoCP}. We now know of three generations of leptons \eqn{eq:lleptons} and quarks \eqn{eq:lquarks}---a good beginning.

A simple test for the completeness of the CKM picture is to ask whether the magnitudes $\abs{V_{ij}}$ are consistent with the hypothesis that the matrix is unitary. Particular attention has been accorded to the first row of the CKM matrix, looking for deviations from the unitarity requirement
\begin{equation}
\mathcal{S}_u \equiv \abs{V_{ud}}^2 + \abs{V_{us}}^2 + \abs{V_{ub}}^2 = 1,
\label{eq:firstrow}
\end{equation}
which would signal new physics. (Because $\abs{V_{ub}}^2 \ll 1$, this is essentially a test of the Cabibbo picture.) For several years, the sum $\mathcal{S}_u$ lay a couple of standard deviations below unity. Recent kaon decay studies have raised the value of $\abs{V_{us}}$, so that $\mathcal{S}_u = 0.9999 \pm 0.0010$~\cite{CKMunitarity}. Ongoing studies of neutron decays should resolve a persistent lifetime puzzle~\cite{Paul:2009md}, and may lead to an improved determination of $\abs{V_{ud}}$.

Immense experimental effort has produced a rich library of information about decays (both common and rare) neutral-particle mixings, and \textsf{CP} violation (in $K$ and $B$ decays)~\cite{CPReview,Browder:2003sq,Hocker:2006xb}. One application of that body of knowledge has been to probe in depth the unitarity of the CKM matrix $\mathsf{V}\mathsf{V}^\dagger = \mathsf{I}$, where $\mathsf{I}$ is the $3 \times 3$ identity, by examining $\sum_i V_{ij}V_{ik}^* = \delta_{jk}$ and $\sum_j V_{ij}V^*_{kj} = \delta_{ik}$. The six vanishing conditions may be represented as triangles in the complex plane, each with an area proportional to $\mathrm{Im}[V_{ij}V_{k\ell}V^*_{i\ell}V^*_{kj}]$, a parametrization-independent measure of \textsf{CP} violation~\cite{Jarlskog:1985ht}. Comprehensive analyses have been carried out over a number of years by the CKM Fitter~\cite{Charles:2004jd} and UTFit~\cite{Bona:2007vi} Collaborations.

\begin{figure}
\centerline{\includegraphics[width=\columnwidth]{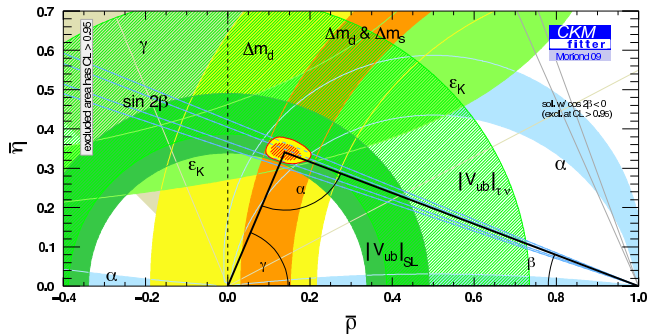}}
\caption{Constraints in the ($\bar{\rho},\bar{\eta}$) plane as of March 2009. The red hashed region shows the global combination at 68\% CL~\cite{Charles:2004jd}. \label{fig:unitri}}
\end{figure}

The most commonly displayed unitarity triangle, shown in  Figure~\ref{fig:unitri}, is constructed from the constraint,
\begin{equation}
V_{ud}V^*_{ub} + V_{cd}V^*_{cb} + V_{td}V^*_{tb} = 0 .
\label{eq:famousut}
\end{equation}
It is conventional to normalize the triangle, dividing the complex vector for each leg by the well-determined $V_{cd}V^*_{cb}$. The vertices of the triangle are then $(0,0)$, $(1,0)$, and $(\bar{\rho},\bar{\eta})$. Among the tests available in this formalism are whether the triangle closes and whether different data sets yield a common vertex, $(\bar{\rho},\bar{\eta})$. The plot in Figure~\ref{fig:unitri}, which is representative of recent work, shows consistency among many experimental constraints. That the imaginary coordinate $\bar{\eta}$ differs from zero shows that the Kobayashi--Maskawa mechanism is at work. A crucial prediction, that $\textsf{CP}$ violation in $K$ physics is small because of flavor suppression but $\textsf{CP}$ violation should be appreciable in $B$ physics, is fulfilled. More detailed analysis shows that the Kobayashi--Maskawa mechanism is the dominant source of $\textsf{CP}$ violation in meson decays. As we have seen in \S\ref{subsec:fcnc}, new physics contributions are extremely small in  $s \leftrightarrow d$, $b \leftrightarrow d$, $s \leftrightarrow b$, and $c \leftrightarrow u$ transitions. For summaries of tests of the CKM paradigm in flavor physics, see~\cite{Pierini:2009zz} for an experimental perspective and a look ahead, and~\cite{Buras:2009us} for a theoretical perspective.

A global fit~\cite{CKMReview}, within the framework of the \textit{three-generation} standard model, yields the following \textit{magnitudes} $\abs{V_{ij}}$ for the CKM matrix elements:
\begin{equation}
{\small
\left(
\begin{array}{ccc}
0.97419 \pm 0.00022 & 0.2257 \pm  0.0010 & 0.00359 \pm 0.00016 \\[4pt]
0.2256 \pm 0.0010 & 0.97334 \pm 0.00023 & 0.0415^{+0.0010}_{-0.0011} \\[4pt]
0.00874^{+0.00026}_{-0.00037} & 0.0407 \pm 0.0010 & 0.999133^{+0.000044}_{-0.000043}
\end{array}
\right).
}
\label{eq:ckmmags}
\end{equation}

The consistency of the CKM picture does not yet exclude a fourth generation of quarks. Direct constraints on $\abs{V_{tb}}$ are consistent with a value near unity, but are not yet terribly restrictive. Global fits to the precision electroweak data allow mixing between the third and fourth families at the level seen between the first and second families~\cite{Chanowitz:2009mz}.

Finally, the robustness of the CKM unitarity triangle does not mean that there is no new physics to be found. The unitarity triangle analysis is mainly sensitive to processes that change flavor by two units. Even in the well-studied rare $K$ and $B$ decays (flavor change by one unit), many examples of new physics that could have passed the unitarity-triangle screen---supersymmetry, little Higgs models with $T$-parity, and warped extra dimensions---could give large departures~\cite{Buras:2009dy}.  New sources of \textsf{CP} violation and FCNC occur in models that do not enforce minimal flavor violation. As we saw in \S\ref{subsec:fcnc}, there is ample space between current bounds and standard-model expectations in many rare decays.  Since the unitarity triangle is described well by the standard model, it will pay to examine \textsf{CP} violation in $b \to s$ transitions and rare decays, where standard-model contributions are small. One specific scenario, involving extra $\mathrm{U(1)}^\prime$ interactions, is presented in~\cite{Barger:2009eq}, and a claimed sign of new physics in $b \to s$ transitions is given in~\cite{Bona:2008jn}.
 
 The ability of the electroweak theory incorporating CKM mixing to account for---and predict---a vast number of observables in flavor physics is highly impressive. We must remember, however, that experiments have validated a framework, not an explanation. Just as the standard model makes no predictions for quark and lepton masses, it has nothing to say about the mixing angles and the Kobayashi--Maskawa phase. These can arise in the electroweak theory, but we don't know how. If quark and lepton masses and mixings are indeed generated by the Higgs mechanism, then (in the words of Veltman) the Higgs boson must know something we do not know~\cite{Veltman:1997nm}. 

 \subsection{Loop-level \label{subsec:loops} }
 We have just recalled some of the ways in which experiment has tested the consequences of the spontaneously broken \ewgg\ gauge theory of the electroweak interactions, and probed with increasing acuity the inferences from earlier experiments on which the electroweak theory was founded. The major predictions for electroweak phenomenology have been confirmed---among them, the existence of neutral-current interactions, the existence and mass scale of the $W^\pm$ and $Z^0$, and the need for the charm quark. The idealizations that shaped the structure of the theory---including the absence of right-handed charged currents and the absence of flavor-changing neutral-current interactions have proved to be exceptionally robust. Only the idealization that the neutrinos are massless has required revision, and that is for many purposes an inessential change.
 
 The electroweak theory is a quantum field theory. Once the elementary interactions have been set by hypothesis or by experimental determinations, we have the opportunity to compute quantum corrections to observables and subject the theory to precise experimental tests. An accessible introduction to the basic techniques can be found in~\cite{Rosner:2001zyA}. The program is straightforward in principle, but very demanding in practice. The mounting precision of experiments has inspired waves of detailed theoretical calculations that are heroic in proportion~\cite{SirlinFest}.
 
 If all the parameters of a theory are known (and the theory is presumed complete), then a measured observable may be compared with the calculated value to test the theory. The electroweak theory has been a work-in-progress over the period when precise measurements became available, because several key parameters have been unknown. 
 
 Before the top quark was discovered in 1995, quantum corrections to electroweak observables gave indications that the weak-isospin partner of $b$ would be much more massive than the other quarks. For example, the 
quantum corrections to the standard-model predictions \eqn{eq:wmass} for 
$M_{W}$ and \eqn{eq:zmass} for $M_{Z}$ arise from different quark loops:
\centerline{ \includegraphics[width=\columnwidth]{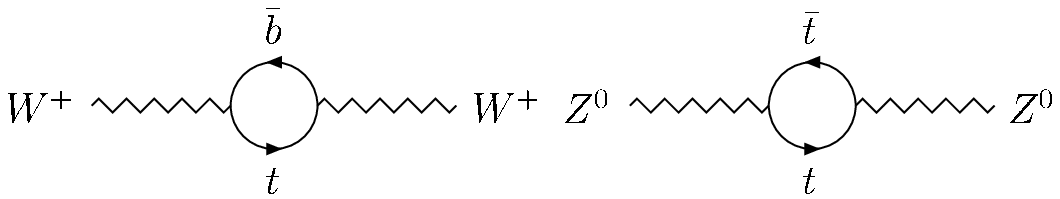}} 
$t\bar{b}$ for $M_{W}$, and $t\bar{t}$ (or $b\bar{b}$) for $M_{Z}$.  
These quantum corrections alter the link  between the $W$- and 
$Z$-boson masses, so that 
\begin{equation}
    M_{W}^{2} = M_{Z}^{2}\left(1 - \sin^{2}\theta_{\mathrm{W}}\right)
    \left(1 + \Delta\rho\right)\; ,
    \label{eq:MWqc}
\end{equation}
where
\begin{equation}
    \Delta\rho \approx \Delta\rho^{(\mathrm{quarks})} = 
    \frac{3G_{\mathrm{F}}m_{t}^{2}}{8\pi^{2}\sqrt{2}} \; .
    \label{eq:drhoq}
\end{equation}
The strong dependence on $m_{t}^{2}$ is characteristic, and 
accounts for the sensitivity of electroweak observables to the top-quark mass.

\begin{figure}[tb]
\centerline{\includegraphics[width=\columnwidth]{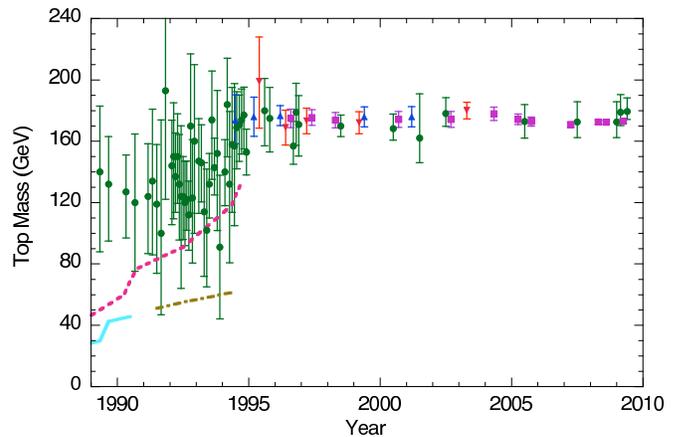}}
\caption{Indirect determinations of the top-quark mass from fits to 
	electroweak observables (green circles) and 95\% confidence-level
	lower bounds on the top-quark 
	mass inferred from direct searches in $e^{+}e^{-}$ annihilations 
	(solid line) and in $\bar{p}p$ collisions, assuming that standard 
	decay modes dominate (broken line).  An indirect lower bound, derived 
	from the $W$-boson width inferred from $\bar{p}p \rightarrow 
	(W\hbox{ or }Z)+\hbox{ anything}$, is shown as the dot-dashed line.  
	A selection of direct measurements of $m_{t}$ by the CDF (blue triangles) and D0
	(inverted red triangles) Collaborations are plotted.  The Tevatron
	average from direct observations is shown as magenta squares.  The most recent indirect determinations are from Refs.~\cite{EWWG,Collaboration:2008ub,Flaecher:2008zq}. The evolution of knowledge of $m_t$ may be traced through the current \textit{Review of Particle Physics}~\cite{Amsler:2008zzb} and previous editions.}
\label{fig:toptimeseries}
\end{figure}

If all other parameters were known, one could choose for any measurement the value of $m_t$ that gave the closest agreement between calculation and experiment, test for consistency among various measurements, and average over different observables, to estimate $m_t$. In practice, the global fits allow for variations in a number of parameters. The top mass favored by simultaneous 
fits to many electroweak observables is shown as a function of time in 
Figure~\ref{fig:toptimeseries}. By the end of 1994, the indirect determinations favored $m_t \approx (175 \pm 25)\gev$, successfully anticipating the masses reported in the discovery papers: $176\pm 8 \pm 10\gev$ for CDF, and $199^{+19}_{-21}\pm 22\gev$ for D0. Today, direct measurements at the Tevatron determine the top-quark mass to a precision of 0.75\%, $m_t = (173.1 \pm 1.3)\gev$~\cite{TopCombo:2009ec}, far more precise than the indirect determinations.

Measurements on and near the $Z^0$ pole by the LEP experiments ALEPH, DELPHI, L3, and OPAL~\cite{Zpole:2005ema} and by the SLD experiment at the Stanford Linear Collider~\cite{Rowson:2001cd} were decisive in testing and refining the electroweak theory~\cite{Altarelli:2004fq}. Global analysis projects that have been distinguished for their thoroughness and continuity include the LEP Electroweak Working Group~\cite{EWWG,Collaboration:2008ub}, incorporating the  \textsf{ZFITTER}~\cite{Bardin:1999yd,Arbuzov:2005ma} and \textsf{TOPAZ0}~\cite{Montagna:1993ai,Montagna:1998kp} codes, and the Particle Data Group~\cite{EWReview}. These have been joined recently by the Tevatron Electroweak Working Group~\cite{TeVEWWG} and the \textsf{Gfitter} initiative~\cite{Flaecher:2008zq}.
 
\begin{figure}
\centerline{\includegraphics[width=0.7\columnwidth]{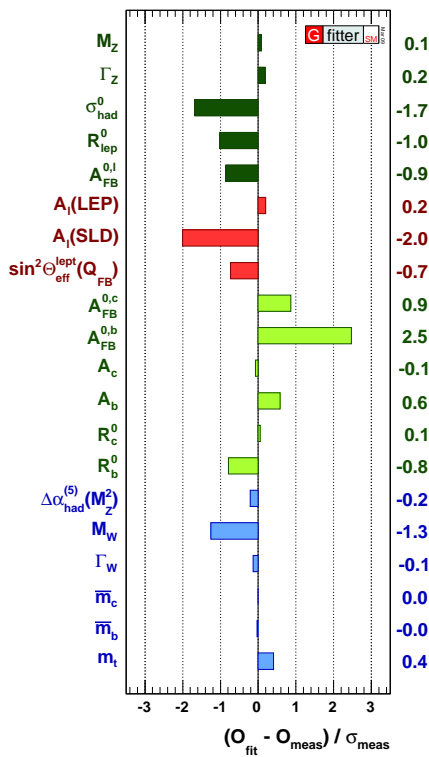}}
\caption{Pull values comparing \textsf{Gfitter} complete fit results with experimental determinations~\cite{Flaecher:2008zq}.}
\label{fig:GfitterPull}
\end{figure}

What has been achieved overall is a comprehensive test of the electroweak theory, as a quantum field theory, at a precision of one part in a thousand for several observables. A representative comparison of best-fit calculations with observations is shown in Figure~\ref{fig:GfitterPull}~\cite{Flaecher:2008zq}, which displays for each observable the difference between fitted and measured values, weighted by the inverse of the experimental standard deviation. [See~\cite{EWWG,Collaboration:2008ub} for the corresponding information from the LEP Electroweak Working Group and~\cite{EWReview} for the Particle Data Group's version.] For only one observable out of twenty---the forward-backward asymmetry in the reaction $e^+ e^- \to b\bar{b}$ on the $Z$ resonance---does the difference exceed two standard deviations. The global fits yield excellent determinations of standard-model parameters, including the weak mixing parameter.

 \subsection{Evidence for Higgs-boson interactions \label{subsec:Hint}}
 An important asset of global fits to many observables is their sensitivity to virtual effects, and thus to parameters that have not been measured directly. The successful inference of the range of top-quark masses is a prime example. Now that $m_t$ is measured at high precision, it becomes a fixed parameter in the global fits, which may probe for the next unknown quantity.
 
\begin{figure}[t!]
	\centerline{\includegraphics[width=\columnwidth]{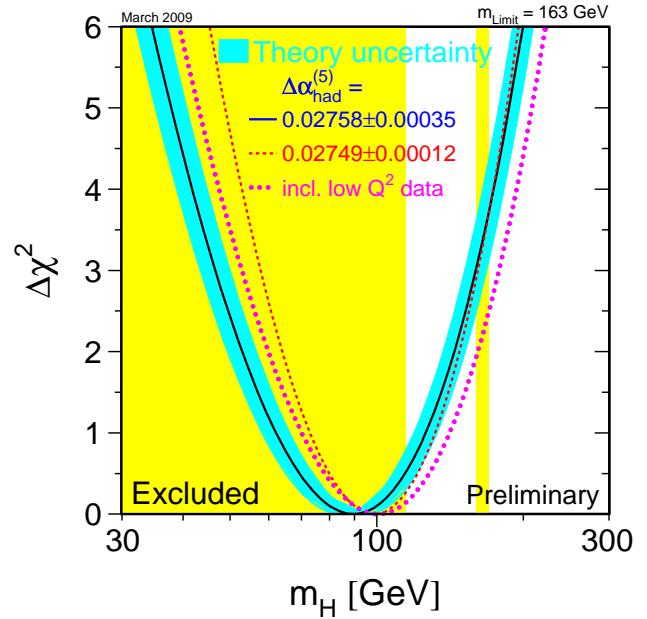}}
	\vspace*{6pt}
	\caption{$\Delta\chi^2 = \chi^2 - \chi_{\mathrm{min}}^2$ from a fit to an ensemble of electroweak measurements as a function of the standard-model Higgs-boson mass. The solid line is the result of the fit.The blue band represents an estimate of the theoretical uncertainty due to missing higher-order corrections. The regions shaded in yellow denote the 95\% CL lower bound on $M_H > 114.4\gev$ from direct searches at LEP~\cite{Barate:2003sz} and the Tevatron exclusion at 95\% CL between 160 and $170\gev$~\cite{Phenomena:2009pt}. The dashed curve shows the sensitivity to a change in the evaluation of $\alpha_{\mathrm{em}}(M_Z^2)$. (From the LEP Electroweak Working Group~\cite{EWWG}.)}
	\protect\label{fig:blueband}
\end{figure}

 Figure~\ref{fig:blueband} shows how the goodness of the LEP Electroweak Working Group's Winter 
2009 global fit depends upon $M_H$. The fit is evidently improved by the inclusion of quantum corrections involving a Higgs boson that has standard-model interactions with the electroweak gauge bosons $W^\pm$ and $Z$. A satisfactory fit does not prove that the standard-model Higgs boson exists, but offers guidance for the search and sets up a consistency check when a putative Higgs boson is observed. The inferred range is consistent with the conditional upper bound, $M_H \lesssim 1\tev$, derived in \S\ref{subsec:tevscale}. It is important to note that, while the global fits give evidence for the effect of the Higgs boson in the vacuum, they do not have any sensitivity to couplings of the Higgs boson to fermions free of the \textit{assumption} that Higgs-Yukawa couplings set the fermion masses.

The precision electroweak measurements on their own argue for $M_H \lesssim 163\gev$, a one-sided 95\% confidence level limit derived from $\Delta\chi^2 = 2.7$ for the blue band in Figure~\ref{fig:blueband}. Imposing the exclusion $M_H > 114.4\gev$ from the LEP searches leads to an upper bound of  $M_H \lesssim 191\gev$~\cite{EWWG}. The Particle Data Group~\cite{EWReview} and \textsf{Gfitter}~\cite{Flaecher:2008zq} analyses lead to similar conclusions.

\begin{figure}
\centerline{\includegraphics[width=\columnwidth]{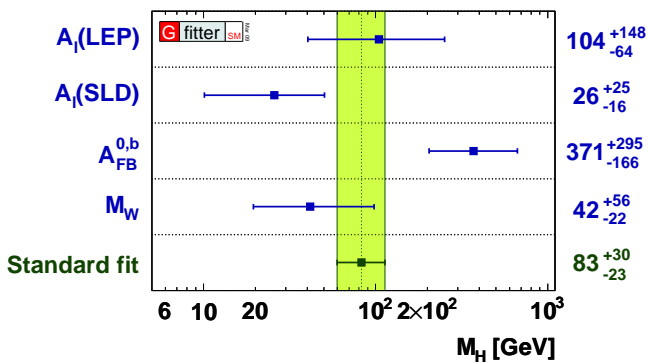}}
\caption{Determination of the Higgs-boson mass excluding all the sensitive observables from the \textsf{Gfitter} standard fit, except for the one given~\cite{Flaecher:2008zq}.}
\label{fig:GfitterHiggsAlt}
\end{figure}

The Higgs-boson masses favored by the global fits of the LEP Electroweak Working Group, $M_H = 90^{+36}_{-27}\gev$~\cite{EWWG}, \textsf{Gfitter,} $83^{+30}_{-23}\gev$~\cite{Flaecher:2008zq}, or Particle Data Group, $70^{+28}_{-22}\gev$~\cite{EWReview}, lie in the region excluded by direct searches at LEP.  Chanowitz~ \cite{Chanowitz:2002cd,Chanowitz:2008ix} has cautioned that the values of $M_H$ preferred by fits to different observables are not entirely consistent. The scatter is illustrated in the case of the \textsf{Gfitter} analysis in Figure~\ref{fig:GfitterHiggsAlt}. In particular, the forward-backward asymmetry in $e^+ e^- \to b\bar{b}$ on the $Z$ resonance ($A^{0,b}_{\mathrm{FB}}$)  is best reproduced with $M_H \approx 400\gev$. This is the observable most discrepant\footnote{For the purpose of this discussion, I set aside the anomalous magnetic moment of the muon~\cite{Bennett:2006fi}, for which the standard-model prediction remains somewhat uncertain. See~\cite{Miller:2007kk}.}, at $\gtrsim 2.5\sigma$, with the overall fits (\cf Figure~\ref{fig:GfitterPull}). Omitting it (on the hypothesis that it is particularly sensitive to new physics) would improve the global fits, but lead to a small Higgs-boson mass that would coexist uncomfortably with the LEP exclusion: the \textsf{Gfitter} best-fit range moves to $61^{+30}_{-26}\gev$. Whether the spread of Higgs-boson masses preferred by different sensitive observables points to physics beyond the standard model or represents insignificant scatter is a tantalizing question.

 \subsection{The weak mixing parameter at low scales}
 The extraordinary precision of measurements on the $Z^0$ pole has given them a decisive weight in our assessment of the electroweak theory. They are, however, blind to new physics that does not directly modify the $Z^0$ properties. A heavy $Z^\prime$ that does not mix appreciably with $Z^0$ is an important example. For this reason, experiments off the $Z^0$ pole, even of lower precision, command our attention---particularly in the search for physics beyond the standard model.
 
 The weak mixing parameter is defined in terms of (running) couplings, 
 \begin{equation}
   \sin^{2}\theta_{\mathrm{W}}(Q) =\frac{\alpha(Q)}{\alpha_{2}(Q)} = 
    \frac{1/\alpha_{2}(Q)}{1/\alpha_{Y}(Q)+1/\alpha_{2}(Q)}\;, \label{eq:xwdef2}
\end{equation} 
 so its value depends on the scale at which it is measured. A familiar illustration occurs in unified theories of the strong, weak, and electromagnetic interactions, which predict the value of the weak mixing parameter at low scales. The prototype is the $\mathrm{SU(5)}$ unified theory~\cite{Georgi:1974sy}: 
 At the unification scale $U \approx  10^{15}\gev$, the running couplings are simply related:
 \begin{equation}
\left.\begin{array}{l}
	1/\alpha_{2} = 1/\alpha_{U}  \\
	1/\alpha_{Y} = \cfrac{5}{3}\cdot 1/\alpha_{U}  \\
	1/\alpha = \cfrac{8}{3}\cdot 1/\alpha_{U}
\end{array}\right\} \;, \label{eq:unicoups}
\end{equation}
 where $\alpha_{U}$ is the common value of the $\mathrm{SU(3)_c}$, $\mathrm{SU(2)}_L$, and $\mathrm{U(1)}$ couplings.
At the $\mathrm{SU(5)}$ unification scale, ${\sin^{2}\theta_{\mathrm{W}}(U) = \cfrac{3}{8}}$. How does $\sin^{2}\theta_{\mathrm{W}}$ evolve? In leading logarithmic approximation and at high scales~\cite{Buras:1977yy}, 
\begin{equation}
    \sin^{2}\theta_{\mathrm{W}}(Q) = \cfrac{3}{8} -\cfrac{5}{8}(b_1 - b_2)\alpha(Q)\log\left(Q^{2}/U^{2}\right)\; ,
\end{equation}
where the beta functions $4\pi b_1 = -4n_g/3 - n_H/10$ and $4\pi b_2 = (22 - 4n_g)/3 - n_H/6$ determine the evolution of $1/\alpha_1$ and $1/\alpha_2$, with $n_g$ the number of fermion generations and $n_H$ the number of Higgs doublets. 
The weak mixing parameter decreases as $Q$ decreases from the unification scale $U$. At the $Z$-boson mass, $ \left.\sin^{2}\theta_{\mathrm{W}}(M_{Z})\right|_{\mathrm{SU(5)}} \approx 0.21$, near (but not near enough) to the measured value, $\left.\sin^{2}\theta_{\mathrm{W}}(M_{Z})\right|_{\mathrm{exp}} = 0.23119 \pm 0.00014$ in the $\overline{\mathrm{\textsc{ms}}}$ scheme~\cite{EWReview}.
 
\begin{figure}
\centerline{\includegraphics[width=\columnwidth]{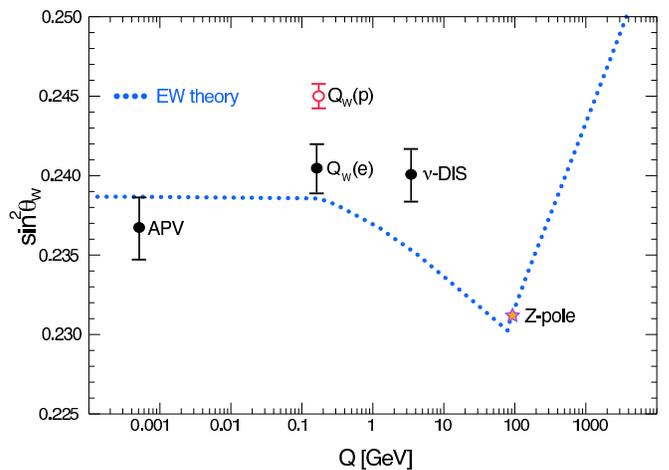}}
\caption{Evolution of the weak mixing parameter $\sin^2\theta_{\mathrm{W}}$ in the $\overline{\mathrm{\textsc{ms}}}$ scheme~\cite{Erler:2004in} (dotted curve). The minimum occurs at $Q=M_W$, where the $\beta$-function for the weak mixing parameter changes sign as the influence of weak-boson loops drops out. The selected data are from atomic parity violation~\cite{Wood:1997zq} (APV), M\o ller scattering~\cite{Anthony:2005pm} ($Q_W(e)$), and deeply inelastic $\nu N$ scallering~\cite{Zeller:2001hh,McFarland:2003jw}. Also indicated (open circle) is the uncertainty projected for the $Q_{\mathrm{weak}}$ experiment~\cite{QWeak}.}
\label{fig:lexw}
\end{figure}

 In the range of scales directly accessible to experiment, the evolution of the weak mixing parameter is predicted within the electroweak theory itself.  The expectations of a higher-order renormalization group analysis~\cite{Erler:2004in} are depicted in Figure~\ref{fig:lexw}.
A detailed comparison with experiment is given in~\cite{EWReview}.  Here are some of the main points. The parity-violating left-right asymmetry observed~\cite{Anthony:2005pm} in polarized M{\o}ller scattering, $e^- e^- \to e^- e^-$, at SLAC establishes the low-energy running of $\sin^2\theta_{\mathrm{W}}$ at more than six standard deviations, and is in reasonable agreement with the prediction at $Q^2 = 0.026\gev^2$. After important improvements in the connection between the measured quantity and $\sin^2\theta_{\mathrm{W}}$, the most telling measurement of atomic parity violation~\cite{Wood:1997zq} agrees with the electroweak theory within about one standard deviation. The $Q_{\mathrm{weak}}$ experiment~\cite{QWeak}, to be mounted at Jefferson Laboratory at the beginning of 2010, aims for a 0.3\% determination of $\sin^2\theta_{\mathrm{W}}$ in parity-violating scattering of polarized electrons on protons at $Q^2 = 0.03\gev^2$.

The  NuTeV experiment at Fermilab determined $\sin^2\theta_{\mathrm{W}}$ by measuring neutral-current and charged-current cross sections for deeply inelastic $\nu N$ and $\bar{\nu}N$ scattering~\cite{Zeller:2001hh,McFarland:2003jw}. Their result, which lies some three standard deviations above the electroweak-theory expectation, has been subjected to intense scrutiny. For the moment, enough ambiguity attends the dependence on fine details of parton distribution functions, the influence of nuclear targets, and various isospin-violating effects that the significance of the NuTeV anomaly is under debate. A catalogue of some ``new physics'' interpretations is given in~\cite{Davidson:2001ji}. Many of these (new $Z^\prime$ gauge bosons~\cite{Chanowitz:2009dz}, leptoquarks, etc.) can be tested at the LHC. New low-energy experiments can test the NuTeV measurement and constrain interpretations. The NuSOnG concept put forward for the Tevatron~\cite{Adams:2008cm} would supplement deeply inelastic $\nu N$ scattering with high-statistics measurements of $\nu e$ and $\bar{\nu}e$ elastic scattering, to test for new physics~\cite{Marciano:2003eq,Gouvea:2006cb} in the neutrino sector.
 
The LEP~2 measurements at energies between the $Z$-pole and the top energy of $209\gev$ were broadly in agreement with standard-model expectations~\cite{Alcaraz:2006mx,EWWG}.
Measurements by the CDF~\cite{CDFZasym,Acosta:2004wq} and D0~\cite{AbaVM:2008xq} experiments of the forward-backward asymmetry in the reaction $\bar{p}p \to (Z,\gamma^*) +\hbox{ anything} \to e^+e^- + \hbox{anything}$ agree with leading-order predictions in the standard model over the range of invariant masses $50\gev \lesssim \mathcal{M}(e^+e^-) \lesssim \hbox{few hundred}\gev$. With the $10\fb^{-1}$ of data expected by the end of Run II, a measurement of the running of 
of $\sin^2\theta_{\mathrm{W}}$ at an interesting level of precision might be achieved before the LHC experiments pronounce on this subject. For a prospectus of low-energy tests of the weak interaction, see~\cite{Erler:2004cx}.

 \subsection{The scale of fermion mass generation \label{subsec:fmassgen}}
\begin{figure}
\centerline{\includegraphics[width=\columnwidth]{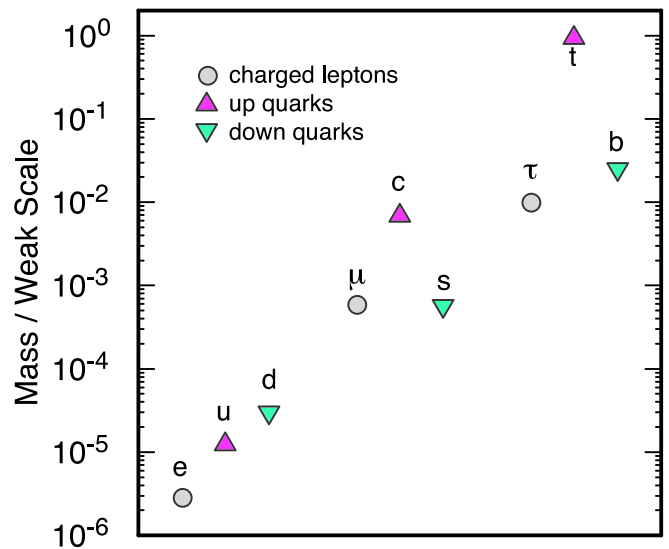}}
\caption{Yukawa couplings $\zeta_i = m_i/(v/\sqrt{2})$ inferred from the masses of the quarks and charged leptons~\cite{Amsler:2008zzb}.}
\label{fig:yuks}
\end{figure}

 It is no exaggeration to say that the origin of the quark and lepton masses is shrouded in mystery. Within the standard electroweak theory, the overall scale of the fermion masses is set by the vacuum expectation value $v/\sqrt{2} \approx 174\gev$ of the Higgs field, but each fermion mass $m_i = \zeta_iv/\sqrt{2}$  involves a distinct Yukawa coupling $\zeta_i$, as we saw in (\ref{eq:Yukterm}). 
 The Yukawa couplings that reproduce the observed quark and charged-lepton 
masses range over many orders of magnitude, from $\zeta_{e} \approx 
3 \times 10^{-6}$ for the electron to $\zeta_{t} \approx 1$ for the 
top quark, as shown in Figure~\ref{fig:yuks}.  
Their origin is unknown.
In an important sense, therefore, \textit{all fermion masses involve physics
beyond the standard model.}
 
 In fact, although the electroweak theory shows how fermion masses might arise, we
  cannot be sure that finding the Higgs
boson, or understanding electroweak symmetry breaking, will bring
clarity about the origin of fermion masses. This is because we do not know that fermion masses are set on the electroweak scale. This point merits closer examination.  

The observation of a nonzero fermion mass ($m_{i} \neq 0$) implies that the electroweak gauge symmetry 
\ewgg\ is broken (\cf \S\ref{sec:ewtheory}), but electroweak symmetry breaking is only a 
necessary, not a sufficient, condition for the generation of fermion 
mass.  In the standard-model framework, some new physics (at an unknown scale) must give rise to the Yukawa couplings. The logical division of labor between a mechanism for electroweak symmetry breaking and an origin of fermion masses is made explicit in the simple technicolor models~\cite{Weinberg:1975gm,Susskind:1978ms} that we shall discuss in \S\ref{subsubsec:techni}). In the sparest versions of such models, electroweak symmetry breaking is driven by a gauge interaction that becomes strongly coupled on the electroweak scale. The gauge bosons acquire masses, but the fermions remain massless. ``Extended technicolor'' models~\cite{Eichten:1979ah,Dimopoulos:1979es,Farhi:1980xs} invoke additional interactions at a much higher scale, of order $100\tev$, to explain the light-quark masses.

Within the framework of the \ewgg\ gauge theory, partial-wave unitarity sets a model-independent upper bound on the energy scale of fermion mass generation~\cite{Appelquist:1987cf}. The strategy is to simply add explicit fermion mass terms to the electroweak Lagrangian, rather than the Yukawa terms of (\ref{eq:Yukterm}).  Explicit Dirac mass terms link the left-handed and right-handed fermions, and thus violate the \ewgg\ gauge symmetry of the electroweak theory. If they persist to arbitrarily high energies, such hard masses destroy the renormalizability of the theory.  On the other hand, it may be overly ambitious to demand that a theory make sense at all energies.  Accordingly, we consider the explicit fermion masses in the framework of an \textit{effective field theory} valid over a finite range of energies, to be supplanted at higher energies by a theory that entails a different set of degrees of freedom~\cite{Georgi:1994qn}.
  
 Because the gauge symmetry is broken in a theory with explicit fermion masses $m_i$, at lowest order in perturbation theory, scattering amplitudes for the production of pairs of longitudinally polarized gauge bosons in fermion-antifermion annihilations grow with c.m.\ energy roughly as $G_\mathrm{F}m_i E_{\mathrm{cm}}$. (In the standard electroweak theory, this behavior is cancelled by the contribution of direct-channel Higgs-boson exchange.) The resulting partial-wave amplitudes saturate partial-wave unitarity for the standard model with a Higgs mechanism at a critical c.m.\ energy~\cite{Appelquist:1987cf, Maltoni:2001dc,Dicus:2004rg},
\begin{equation}
\sqrt{s_i} \simeq \frac{4\pi\sqrt{2}}{\sqrt{3\eta_i} \, G_\mathrm{F} m_i} = 
                \frac{8 \pi v^2}{\sqrt{3\eta_i} \, m_i}\;,
\label{unitaritybound}
\end{equation}
where  $\eta_i=1 (3)$ for leptons (quarks). 
As usual, the parameter $v$ sets the scale of electroweak symmetry breaking. If the electron mass were hard, the critical energy would be $\sqrt{s_e} \approx 1.7 \times 10^9\gev$; the corresponding energy for the top quark is $\sqrt{s_t} \approx 3\tev$. The fact that a hard electron mass would only imply a saturation of partial-wave unitarity at a prodigiously high energy means that while the behavior of $\sigma(e^+ e^- \to W^+ W^-)$ shown in Figure~\ref{fig:LEPgc} validates the gauge symmetry of the electroweak theory, it does not establish that the theory is renormalizable~\cite{Christensen:2005hm,Appelquist:1987cf}.

\section{THE AGENT OF ELECTROWEAK SYMMETRY BREAKING \label{sec:agent}}
\subsection{The significance of the 1-TeV scale \label{subsec:tevscale}}
The electroweak theory does not give a precise prediction for the mass of the Higgs boson, but a thought experiment leads through a unitarity argument~\cite{Lee:1977eg} to a conditional upper bound on the Higgs-boson mass that sets a key target for experiment. 

Consider two-body collisions among $W^\pm$. $Z^0$, and $H$.
It is straightforward to compute the scattering
amplitudes ${\cal M}$ at high energies, and to make
a partial-wave decomposition, according to ${\cal M}(s,t)=16\pi\sum_J(2J+1)a_J(s)P_J(\cos{\theta})$.
 Most channels ``decouple,'' in the sense 
that partial-wave amplitudes are small at all energies (except very
near particle poles, or at exponentially large energies), for
any value of the Higgs boson mass $M_H$. Four neutral channels are interesting:
\begin{equation}
\begin{array}{cccc}
W_0^+W_0^-\quad & {\displaystyle \frac{Z_0Z_0}{\sqrt{2}}}\quad & {\displaystyle \frac{HH}{\sqrt{2}}}\quad & HZ_0 \; ,
\end{array}
\end{equation}
where the subscript $0$ denotes the longitudinal polarization
states, and the factors of $\sqrt{2}$ account for identical particle
statistics. For these, the $s$-wave amplitudes are all asymptotically
constant (\ie, well-behaved) and  
proportional to $G_{\mathrm{F}}M_H^2.$ In the high-energy 
limit $s \gg M_H^2, M_W^2, M_Z^2$,
\begin{equation}
(a_0)\to\frac{-G_{\mathrm{F}} M_H^2}{4\pi\sqrt{2}}\cdot \left[
\begin{array}{cccc} 1 & 1/\sqrt{8} & 1/\sqrt{8} & 0 \\
      1/\sqrt{8} & 3/4 & 1/4 & 0 \\
      1/\sqrt{8} & 1/4 & 3/4 & 0 \\
      0 & 0 & 0 & 1/2 \end{array} \right] \; .
\end{equation} 
Requiring that the largest eigenvalue respect the 
partial-wave unitarity condition $\abs{a_0}\le 1$ yields
\begin{equation}
	M_H \le \left(\frac{8\pi\sqrt{2}}{3G_{\mathrm{F}}}\right)^{1/2} \approx 1\tev
	\label{eq:higgsbnd}
\end{equation}
as a condition for perturbative unitarity.

If the Higgs-boson mass respects the bound \eqn{eq:higgsbnd}, weak interactions remain weak at all
energies, and perturbation theory is everywhere reliable. If the Higgs-boson mass exceeds $1\tev$, perturbation theory breaks down, as weak interactions among $W^\pm$, $Z$, and $H$ become strong on the \onetev. This means that (within the standard model) the features familiar in strong-interaction physics at GeV energies would characterize electroweak-boson interactions at
TeV energies. More generally, the implication is that something new---a Higgs boson, strong scattering, or other new physics---is to be found in electroweak interactions at energies not much larger than $1\tev$.

Tighter constraints---in the form of upper and lower bounds on the mass of the Higgs boson---follow from the demand that the electroweak theory be a consistent (and complete) quantum field theory, up to a specified energy scale $\Lambda$.\footnote{The substantial literature on this topic may be traced from the state-of-the-art papers cited here.} For a light Higgs boson, the $t\bar{t}H$ Yukawa coupling introduces quantum corrections that may destabilize the Higgs potential \eqn{eq:scalarpot} so that the electroweak vacuum state characterized by \eqn{eq:vevis} is no longer the state of minimum energy. The perturbative analysis is explained carefully in~\cite{Einhorn:2007rv}. For a specified value of the top-quark mass, the requirement that the broken-symmetry vacuum of the electroweak theory be the absolute minimum of the (radiatively corrected) Higgs potential gives a lower bound on the Higgs-boson mass. For a cutoff $\Lambda = 1\tev$, the lower bound is~\cite{Casas:1996aq}
\begin{equation}
\left.M_H\right|_{\Lambda = 1 \mathrm{TeV}} \gtrsim  50.8\gev + 0.64 (m_t - 173.1\gev) ,
\label{eq:MHlow}
\end{equation}
already surpassed by searches at LEP, while for $\Lambda = M_{\mathrm{Planck}}$, the lower bound rises to 
\begin{equation}
\left. M_H \right|_{\Lambda = M_{\mathrm{Planck}}} \gtrsim 134\gev.
\label{eq:MHlowP}
\end{equation}

Only noninteracting, or \textit{trivial,} scalar field theories make sense on all energy scales. With restrictions, such theories can make sense up to a specified scale $\Lambda$ at which new physics comes into play. By analyzing the $Q^2$-evolution of the running quartic coupling in \eqn{eq:scalarpot}, it is possible to establish an upper bound on the coupling, and hence on the Higgs-boson mass, at some reasonable scale accessible to experiment. A two-loop analysis leads to the bounds~\cite{Hambye:1996wb}
\begin{eqnarray}
\left. M_H \right|_{\Lambda = M_{\mathrm{Planck}}} & \lesssim 180\gev ; \label{eq:MHupP} \\
\left.M_H\right|_{\Lambda = 1 \mathrm{TeV}} & \lesssim 700\gev . \label{MHup}
\end{eqnarray}

The electroweak theory could in principle be self-consistent up to very high energies, provided that the Higgs-boson mass lies in the interval $134\gev \lesssim M_H \lesssim 180\gev$. If $M_H$ lies outside this band, new physics will intervene at energies below the Planck (or unification) scale.

It is of considerable interest to use the techniques of lattice field theory to explore nonperturbative aspects of Higgs physics. What has been learned so far can be traced from~\cite{Fodor:2007fn,Gerhold:2009ub}.

An informative perspective on the lower bound \eqn{eq:MHlowP} can be gained by relaxing the requirement that the electroweak vacuum correspond to the absolute minimum of the Higgs potential. It is consistent with observations for the ground state of the electroweak theory to be a \textit{false} (metastable) vacuum that has survived quantum fluctuations until now. The relevant constraint is then that the mean time to tunnel from our electroweak vacuum to a deeper vacuum exceeds the age of the Universe, about $13.7\hbox{ Gyr}$~\cite{Komatsu:2008hk}.

\begin{figure}[tb]
\centerline{\includegraphics[width=\columnwidth]{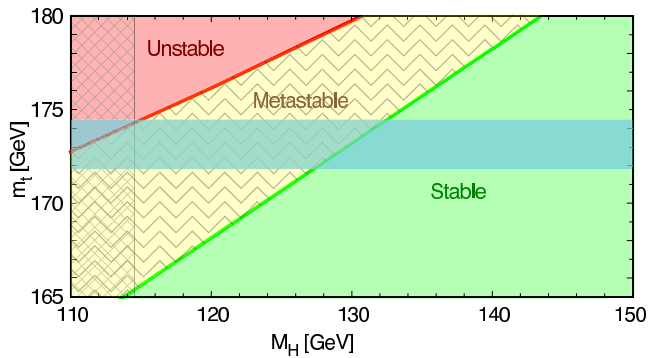}}
\caption{Metastability region of the
standard-model vacuum in the $(M_H,m_t)$ plane~\cite{Isidori:2007vm}. The hatched region at left indicates the LEP lower bound, $M_H > 114.4\gev$. The horizontal band shows the measured top-quark mass, $m_t = (173.1 \pm 1.3)\gev$~\cite{TopCombo:2009ec}.}
\label{fig:onthebrink}
\end{figure}

Figure~\ref{fig:onthebrink} shows the $(M_H,m_t)$ regions in which the standard-model vacuum is stable, acceptably long-lived, or too short-lived, as inferred from a renormalization-group-improved one-loop calculation of the tunneling probability at zero temperature~\cite{Isidori:2007vm}.
Present constraints on $(M_H,m_t)$ suggest that we do not live in the unstable vacuum that would mandate new physics below the Planck scale. At the lowest permissible Higgs-boson mass, this conclusion holds only at 68\% CL.

\subsection{Experimental constraints on the Higgs boson \label{subsec:excon}}
We have seen in our discussion of evidence for the virtual influence of the Higgs boson in \S\ref{subsec:Hint} that global fits, made within the framework of the standard electroweak theory, favor a light Higgs boson, and exhibit some tension with direct searches. The LEP experiments, which focused on the $e^+ e^- \to HZ^0$ channel, set a lower bound on the standard-model Higgs-boson mass of $M_H > 114.4\gev$ {at 95\% CL}~\cite{Barate:2003sz,Kado:2002er}. The Tevatron experiments CDF and D0 also search for the standard-model Higgs boson, examining a variety of production channels and decay modes appropriate to different Higgs-boson masses. The most recent combined result excludes the range $160\gev < M_H < 170\gev$  at 95\% CL~\cite{Bernardi:2008ee,Phenomena:2009pt}.
See~\cite{HiggsPDG08} for an overview of past searches.

\begin{figure}[tb]
\centerline{\includegraphics[width=\columnwidth]{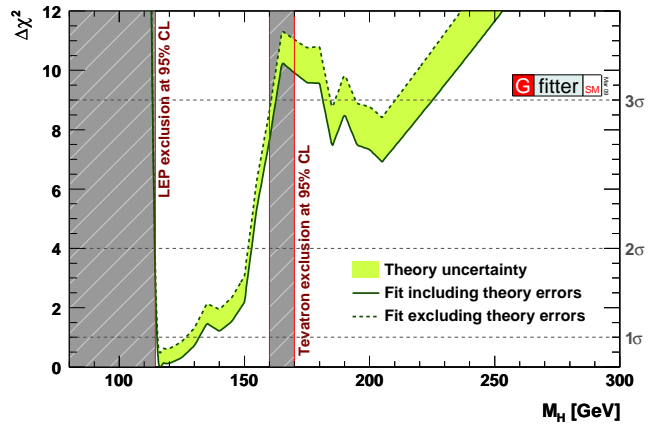}}
\caption{$\Delta\chi^2$ as a function of the Higgs-boson mass for the \textsf{Gfitter} complete fit, taking account of direct searches at LEP and the Tevatron. The solid (dashed) line gives the results when including (ignoring) theoretical errors. The minimum $\Delta\chi^2$ of the fit including theoretical errors is used for both curves to obtain the offset-corrected $\Delta\chi^2$~\cite{Flaecher:2008zq}.}
\label{fig:GfitterHiggsSearch}
\end{figure}

The disjoint exclusion regions from LEP and the Tevatron make it somewhat complicated to specify the remaining mass ranges favored for the standard-model Higgs boson. A useful example is shown in Figure~\ref{fig:GfitterHiggsSearch}~\cite{Flaecher:2008zq}.
In the \textsf{Gfitter} analysis, at $2\sigma$-significance ($\approx$ 95\% CL), the standard-model Higgs-boson mass must lie in the interval $113.8\gev < M_H < 152.5\gev$.

The standard electroweak theory gives an excellent account of many pieces of data over a wide range of energies, and its main elements can be stated compactly. Nevertheless, it leaves too many gaps in our understanding for it to be considered a complete theory (\cf \S~\ref{sec:SMincomplete}). We therefore have reason to consider extensions to the standard model, for which the standard-model fits to the electroweak measurements do not apply. Accordingly, healthy skepticism dictates that we regard the inferred constraints on $M_H$ as a potential test of the standard model, not as rigid boundaries on where the agent of electroweak symmetry breaking must show itself.

Supersymmetric extensions of the electroweak theory entail considerable model-dependence, but yield high-quality fits to the precision data~\cite{Erler:1998ur,Heinemeyer:2007bw,AbdusSalam:2009qd}. Bounds inferred from searches for the lightest \textsf{CP}-even Higgs boson $h$ of the minimal supersymmetric standard model are somewhat less restrictive than for the standard-model Higgs boson. The tension between fits that prefer light masses and direct searches that disfavor a light Higgs boson is not present in the supersymmetric world. On the other hand, in its simplest form, the minimal supersymmetric standard model would be challenged if $M_h$ exceeded about $135\gev$. A thorough discussion appears in \S7.1 of~\cite{Martin:1997ns}. A recent 25-parameter fit to the ``phenomenological minimal supersymmetric standard model'' concludes that $117\gev \lesssim M_h \lesssim 129\gev$~\cite{AbdusSalam:2009qd}.

If new strong dynamics---rather than a perturbatively coupled elementary scalar---hides the electroweak symmetry, then the mass of the composite stand-in for the Higgs boson can range up to several hundred GeV. The same is true for standard-model fits that allow an extra generation of quarks and leptons~\cite{Kribs:2007nz,Novikov:2009kc}.

It is prudent that we plan to search for the agent of electroweak symmetry breaking over the entire mass range allowed by general arguments, and this is what the LHC experiments will do. As an illustration, we next consider some elements of a broad search for the standard-model Higgs boson. This is a point of departure for more exotic searches. The search for the Higgs boson is now the province of the proton accelerators. The 2-TeV proton-antiproton Tevatron Collider is operating now, its integrated luminosity having surpassed $6\fb^{-1}$, and the 14-TeV Large Hadron Collider at CERN will provide high-luminosity proton-proton collisions beginning in 2009.

\subsection{Search for the standard-model Higgs boson \label{subsec:Hsearch}}
The search for the Higgs boson has been a principal goal of particle physics for many years, so theorists and experimentalists have explored search strategies in great detail. The techniques in use at the Tevatron may traced from~\cite{Phenomena:2009pt}, while the protocols foreseen for experiments at the Large Hadron Collider are detailed in the ATLAS~\cite{Aad:2009wy} and CMS~\cite{Ball:2007zza} performance documents.



Because the standard-model Higgs boson gives mass to the fermions and weak gauge bosons, it decays preferentially into the most massive states that are kinematically accessible.  Decays $H \to f\bar{f}$ into fermion pairs, where $f$ occurs in $N_{\mathrm{c}}$ colors, proceed at a rate
\begin{equation}
	\Gamma(H \to f\bar{f}) = \frac{G_{\mathrm{F}}m_{f}^{2}M_{H}}{4\pi\sqrt{2}} 
	\cdot N_{\mathrm{c}} \cdot \left( 1 - \frac{4m_{f}^{2}}{M_{H}^{2}} 
	\right)^{3/2} \; ,
	\label{eq:Higgsff}
\end{equation}
which is proportional to $N_{\mathrm{c}} m_f^2 M_{H}$ as the Higgs-boson mass becomes large.
The partial width for decay into a $W^{+}W^{-}$ pair is
\begin{equation}
	\Gamma(H \to W^{+}W^{-}) = \frac{G_{\mathrm{F}}M_{H}^{3}}{32\pi\sqrt{2}} 
	(1 - x)^{1/2} (4 -4x +3x^{2}) \; ,
	\label{eq:HiggsWW}
\end{equation}
where $x \equiv 4M_{W}^{2}/M_{H}^{2}$.  Similarly, the partial width 
for decay into a pair of $Z^{0}$ bosons is 
\begin{equation}
	\Gamma(H \to Z^{0}Z^{0}) = \frac{G_{\mathrm{F}}M_{H}^{3}}{64\pi\sqrt{2}} 
	(1 - x^{\prime})^{1/2} (4 -4x^{\prime} +3x^{\prime 2}) \; ,
	\label{eq:HiggsZZ}
\end{equation}
where $x^{\prime} \equiv 4M_{Z}^{2}/M_{H}^{2}$.  The rates for decays into 
weak-boson pairs are asymptotically proportional to $M_{H}^{3}$ and 
$\cfrac{1}{2}M_{H}^{3}$, respectively.  In the final factors of \eqn{eq:HiggsWW} 
and \eqn{eq:HiggsZZ}, $2x^{2}$ and $2x^{\prime 2}$, respectively, 
arise from decays into transversely polarized gauge bosons.  The 
dominant decays for large $M_{H}$ are into pairs of longitudinally 
polarized weak bosons.  

\begin{figure}[tb]
	\centerline{\includegraphics[width=\columnwidth]{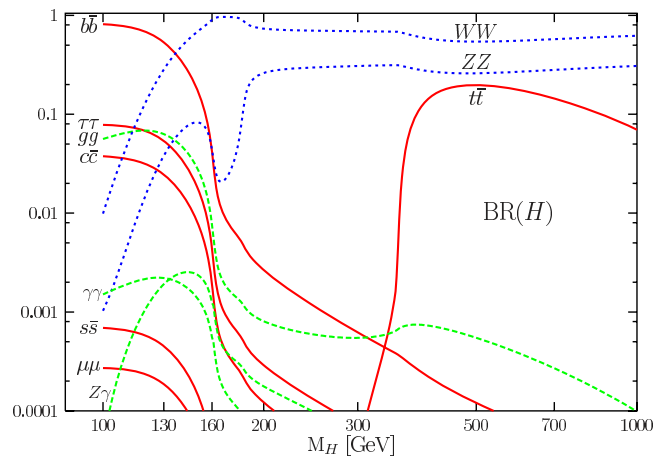}}
	\vspace*{6pt}
	\caption{Branching fractions for prominent decay modes of the standard-model Higgs boson, from~\cite{Djouadi:2005gi}.}
	\protect\label{fig:LHdk}
\end{figure}
Branching fractions for decay modes that may hold promise for the 
detection of a Higgs boson are displayed in Figure 
\ref{fig:LHdk}.  In addition to the $f\bar{f}$ and $VV$ modes that 
arise at tree level, the plot includes the $\gamma\gamma$, $Z\gamma$, and two-gluon modes that 
proceed through loop diagrams. 

\begin{figure}[tb]
	\centerline{\includegraphics[width=\columnwidth]{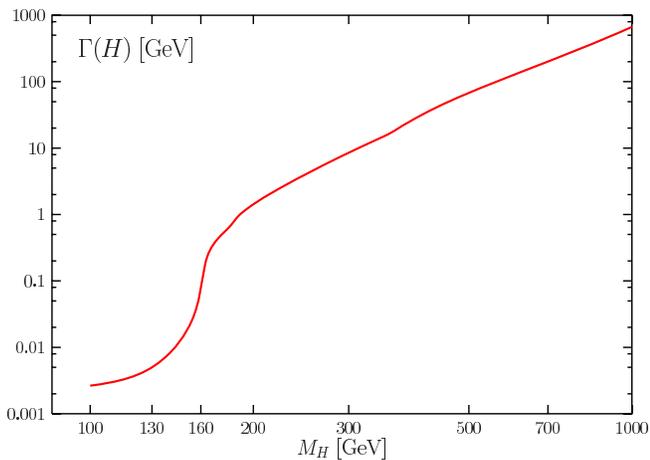}}
	\vspace*{6pt}
	\caption{Total width of the standard-model Higgs boson vs.\ mass, from~\cite{Djouadi:2005gi}.}
	\protect\label{fig:Htot}
\end{figure}

The Higgs-boson total width 
is plotted as a function of $M_{H}$ in Figure~\ref{fig:Htot}. Below the $W$-pair threshold, the standard-model Higgs boson is rather narrow,
with $\Gamma(H \to \mathrm{all}) \lesssim 1\gev$.  Far above the threshold for decay into gauge-boson pairs, the total width is proportional to $M_{H}^{3}$.  As its mass increases toward $1\tev$, the Higgs boson becomes highly unstable, with a perturbative width approaching its mass.  It would therefore be observed as an enhanced rate, rather than a distinct resonance. 

\begin{figure}[tb]
\begin{center}
\includegraphics[width=\columnwidth]{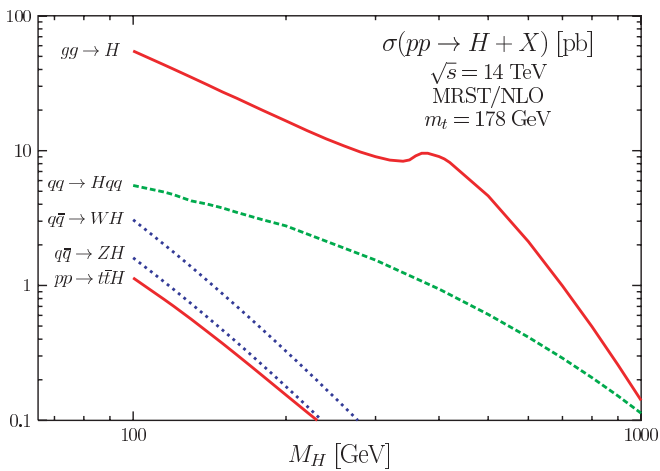}
\caption{Higgs-boson production cross sections in  $pp$ collisions at $\sqrt{s} = 14\tev$, computed at next-to-leading order using the MRST parton distributions~\cite{Martin:2002dr}; from~\cite{Djouadi:2005gi} \label{fig:TeVLHCH}}
\end{center}
\end{figure}

Cross sections for the principal reactions to be studied at the LHC are shown in Figure~\ref{fig:TeVLHCH}.
The largest cross section for Higgs production at both the LHC and the Tevatron occurs in the reaction $p^{\pm}p \to H + \hbox{anything}$, which proceeds by gluon fusion through heavy-quark loops. [The shoulder in that cross section near $M_H = 400\gev$ reflects the behavior of the top-quark loop.] A fourth generation of heavy quarks would raise the $gg \to H$ rate significantly, increasing the sensitivity of searches at the Tevatron and LHC.

For small Higgs-boson masses, the dominant decay is into $b\bar{b}$ pairs, but the reaction $p^{\pm}p \to H + \hbox{anything}$ followed by the decay $H \to b\bar{b}$ is swamped by QCD production of $b\bar{b}$ pairs. Consequently, experiments must rely on rare decay modes ($\tau^+\tau^-$ or $\gamma\gamma$, for example) with lower backgrounds, or resort to different production mechanisms for which specific reaction topologies reduce backgrounds.  Accordingly, the production of Higgs bosons in association with electroweak gauge bosons is receiving close scrutiny at the Tevatron. The rare $\gamma\gamma$ channel is seen as an important target for LHC experiments, if the Higgs boson is light. Fine resolution of the electromagnetic calorimenters is a prerequisite to overcoming standard-model backgrounds. At higher masses, the Tevatron experiments have exploited good sensitivity to the $gg \to H \to W^+W^-$ reaction chain to set their exclusion limits~\cite{Phenomena:2009pt}.

At the LHC, the multipurpose CMS and ATLAS detectors  will make a comprehensive exploration of the Fermi scale, with high sensitivity to the standard-model Higgs boson reaching to $1\tev$. Current projections suggest that a few tens of$\fb^{-1}$ will suffice for a robust discovery~\cite{Aad:2009wy,Ball:2007zza}.

Once the Higgs boson is found, it will be of great interest to map its decay pattern, in order to characterize the mechanism of electroweak symmetry breaking. It is by no means guaranteed that the same agent hides electroweak symmetry and generates fermion mass. In the following \S\ref{subsubsec:qcdhides}, we shall see how chiral symmetry breaking in QCD could hide the electroweak symmetry without generating fermion masses.  Indeed, many extensions to the standard model significantly alter the decay pattern of the Higgs boson.  In supersymmetric models, five Higgs bosons are expected, and the branching fractions of the lightest one may be very different from those presented in Figure~\ref{fig:LHdk}~\cite{Djouadi:2005gj}. 

Precise determinations of Higgs-boson couplings is one of the strengths of the projected International Linear Collider~\cite{Murayama:1996ec,Dawson:2004xz}, but the LHC will supply crucial clues to the origin of fermion masses. For example, a Higgs-boson discovery in gluon fusion ($gg \to H$), signalled by the large production rate, would argue for a nonzero coupling of the Higgs boson to top quarks---an important qualitative conclusion. In time, and by comparing with other production and decay channels,  it should be possible to constrain the $Ht\bar{t}$ coupling. With the LHC's large data sets, it is plausible that Higgs-boson couplings can eventually be measured at levels that test the standard model and provide interesting constraints on extensions to the electroweak theory.

\subsection{Alternatives to the Higgs mechanism \label{subsec:alt}}

\subsubsection{How QCD would hide electroweak symmetry \label{subsubsec:qcdhides}}
An analogy between electroweak symmetry breaking and the superconducting phase transition led to the insight of the Higgs mechanism. The macroscopic order parameter of the Ginzburg-Landau phenomenology, which corresponds to the wave function of superconducting 
charge carriers, acquires a nonzero vacuum expectation value in the 
superconducting state. Within a superconductor, the photon acquires a mass $M_\gamma = \hbar/\lambda_{\mathrm{L}}$, where the London penetration depth, $\lambda_{\mathrm{L}}$, characterizes the exclusion of magnetic flux by the Meissner effect. In the particle-physics counterpart, auxiliary scalars introduced to hide the electroweak symmetry pick up a nonzero vacuum expectation value that gives rise to masses for the $W^\pm$ and $Z^0$.

A deeper look at superconductivity reveals an example of a gauge-symmetry-breaking mechanism that does not rely on introducing an \textit{ad hoc} order parameter. In the microscopic Bardeen-Cooper-Schrieffer theory \cite{Bardeen:1957kj}, the order parameter arises dynamically, through 
the formation of correlated states of elementary fermions, the Cooper pairs of 
electrons. Part of the beauty of the BCS theory is that the only new ingredient required is insight.
The elementary fermions---electrons---and gauge interactions---QED---needed to generate the correlated pairs are already present in the case of superconductivity. This suggests that the electroweak symmetry might also be broken dynamically, without the need to introduce scalar fields. Indeed, quantum chromodynamics can be the source of electroweak symmetry breaking.

Consider an \smgg\ theory of massless up and 
down quarks. Because the strong interaction is strong and the electroweak 
interaction is feeble, we may treat the \ewgg\
interaction as a perturbation. For vanishing quark masses, QCD displays an exact 
$\mathrm{SU(2)_L\otimes SU(2)_R}$ chiral symmetry. At an energy scale 
$\sim\Lambda_{\mathrm{QCD}},$ the strong interactions become strong and quark 
condensates of the form
\begin{equation}
\langle \bar{q}q \rangle \equiv \langle\bar{u}u + \bar{d}d\rangle
\label{eq:qcond}
\end{equation}
appear. The chiral symmetry is spontaneously broken
to the familiar flavor symmetry, isospin:
\begin{equation}
	\mathrm{SU(2)_L\otimes SU(2)_R \to SU(2)_V}\;\; ,
\end{equation}
because the left-handed and right-handed quarks communicate through $\langle\bar{q}q\rangle = \langle\bar{q}_{\mathrm{R}}q_{\mathrm{L}}+\bar{q}_{\mathrm{L}}q_{\mathrm{R}}\rangle$.
 Three Goldstone bosons appear, one for 
each broken generator of the original chiral invariance. These were 
identified by Nambu~\cite{Nambu:1960xd} as three massless pions.

The broken generators are three axial currents whose couplings to pions are 
measured by the pion decay constant $f_\pi \approx 92.4\mev$~\cite{Amsler:2008zzb}, which is measured by the charged-pion lifetime. When we turn on the 
 electroweak interaction, the electroweak gauge symmetry is broken because the left-handed and right-handed quarks, now coupled through the $\langle\bar{q}q\rangle$ condensate, transform differently under \ewgg\ gauge transformations. The electroweak
bosons couple to the axial currents and acquire masses of order $\sim 
gf_\pi$. The mass-squared matrix,
\begin{equation}
	\mathcal{M}^{2} = \left(
		\begin{array}{cccc}
		g^{2} & 0 & 0 & 0  \\
		0 & g^{2} & 0 & 0  \\
		0 & 0 & g^{2} & gg^{\prime}  \\
		0 & 0 & gg^{\prime} & g^{\prime2}
	\end{array}
		 \right) \frac{f_{\pi}^{2}}{4} \; ,
	\label{eq:csbm2}
\end{equation}
(where the rows and columns correspond to $b_1$, $b_2$, $b_{3}$, 
and $\mathcal{A}$) has the same structure as the mass-squared matrix 
for gauge bosons in the standard electroweak theory.  

Diagonalizing 
the matrix (\ref{eq:csbm2}), we find that the photon, corresponding as in the standard model to the combination $A = (g\mathcal{A} + g^{\prime}b_3)/\sqrt{g^2 + g^{\prime 2}}$, emerges massless.
Two charged gauge bosons, $W^{\pm} = (b_1 \mp ib_2)/\sqrt{2}$, acquire mass-squared
$M_{W}^{2} = g^{2}f_{\pi}^{2}/4$, and a neutral gauge boson $Z = (-g^{\prime}\mathcal{A} + gb_3)/ \sqrt{g^2 + g^{\prime 2}}$ obtains $M_{Z}^{2} = 
(g^{2}+g^{\prime2})f_{\pi}^{2}/4$. The ratio,
\begin{equation}
	{M_{Z}^{2}}/{M_{W}^{2}} ={(g^{2}+g^{\prime2})}/{g^{2}} = 
	{1}/{\cos^{2}\theta_{\mathrm{W}}}\; ,
	\label{eq:wzrat}
\end{equation}
where $\theta_{\mathrm{W}}$ is the weak mixing parameter, reproduces the standard-model result.
The would-be massless pions disappear from the physical spectrum, 
becoming the longitudinal components of the weak gauge bosons. 
Here the symmetry breaking is dynamical and automatic; it can be traced, through spontaneous chiral symmetry breaking and confinement, to the asymptotic freedom of QCD. 

Electroweak symmetry breaking determined by pre-existing dynamics stands in
contrast to the standard electroweak theory, in which spontaneous symmetry breaking results from the
{\it ad hoc} choice of $\mu^2 < 0$  for the coefficient of the quadratic
term in the Higgs potential.
Despite the structural similarity to the standard model, the chiral symmetry breaking of QCD does not yield a satisfactory theory of the weak interactions. The masses acquired by the 
intermediate bosons are $2\,500$ times smaller than required for a successful 
low-energy phenomenology; the $W$-boson mass is only~\cite{Weinstein:1973gj} $M_W\approx 30\mev$, because its scale is set by $f_{\pi}$. Moreover, QCD does not give masses to the fermions: the up and down quark and the electron all remain massless. We have already remarked on the logical separation between electroweak symmetry breaking and fermion mass generation in \S\ref{subsec:fmassgen}.

\subsubsection{If no Higgs mechanism shaped the world \ldots}

Having recalled that QCD induces the breaking of electroweak symmetry through the formation of $\langle\bar{q}q\rangle$ condensates, it is worth pausing for a moment to ask how different the world would have been, without a Higgs mechanism or a substitute on the real-world electroweak scale~\cite{Quigg:2009xr}. Eliminating the Higgs mechanism does not alter the strong interaction, so QCD would still confine colored objects into hadrons. In particular, the gross features of nucleons derived from QCD---such as nucleon masses---would be little changed if the up and down quark masses were set to zero. 

In the real world, the small $m_d > m_u$ mass difference overcomes the electromagnetic mass shift that would render the proton heavier than the neutron, and results in $M_n - M_p \approx 1.293\mev$, so that the neutron is unstable to $\beta$ decay and the proton is the lightest nucleus. This contribution is absent if the quarks are massless.

However, the fact that electroweak symmetry is broken on the QCD scale means that the strength of the ``weak'' interactions would be similar to the strength of the strong nuclear force. The analogue of the Fermi constant, $G_{\mathrm{F}}$, is enhanced by nearly seven orders of magnitude. This has many consequences, including the acceleration of $\beta$-decay rates and the amplification of \textit{weak-interaction} mass shifts that tend to make the neutron outweigh the proton. Because the theory lacks a Higgs boson, scattering among weak bosons becomes strongly coupled on the hadronic scale, following the analysis we reviewed in \S\ref{subsec:tevscale}. 

Should the proton be stable, or compound nuclei be produced and survive to late times in this alternate universe, the infinitesimal electron mass would compromise the integrity of matter. The Bohr radius of a would-be atom would be macroscopic (if not infinite), valence bonding would have no meaning, and stable structures would not form. In seeking the agent of electroweak symmetry breaking, we hope to learn why the everyday world is as we find it: why atoms and chemistry and stable structures can exist.

\subsubsection{Dynamical symmetry breaking \label{subsubsec:techni}}
The observation that QCD dynamically breaks electroweak symmetry (but at too low a scale) inspired the invention of analogous no-Higgs theories in which dynamical symmetry breaking is accomplished by  the formation of a condensate of new fermions subject either to QCD itself, or to a new, asymptotically free, vectorial gauge interaction (often called
technicolor) that becomes strongly coupled at the TeV scale. 

Within QCD, hypothetical exotic (color $\mathbf{6}, \mathbf{8}, \mathbf{10,}$ \ldots) quarks would interact more strongly through than the normal color triplets, so the chiral-symmetry breaking in exotic quark sectors would occur at much larger mass scales than the standard chiral-symmetry breaking we have just reviewed. If those mass scales were sufficiently high, exotic-quark condensates could break $\ewgg \to \emgg$ dynamically and yield phenomenologically viable $W^\pm$ and $Z^0$ masses~\cite{Marciano:1980zf}.  No exotic quarks have yet been detected, either by direct observation or in the evolution of the strong coupling constant, $\alpha_s$~\cite{Bethke:2006ac}.

Technicolor theories posit new \textit{technifermions} that are subject to a new technicolor interaction.
 The technifermion condensates that
dynamically break the electroweak symmetry produce masses for the $W^\pm$ and $Z^0$
bosons. Choosing the scale on which the technicolor interaction becomes strong so that the technipion decay constant is given by $F^2_\pi = G_\mathrm{F}\sqrt{2}$ reproduces the gauge-boson masses of the standard electroweak theory.  Technicolor shows how the generation of intermediate boson masses 
could arise without fundamental scalars or unnatural adjustments of 
parameters.  By replacing the elementary Higgs boson with an object that is composite on the electroweak scale, it also offers an elegant solution to the naturalness 
problem of the standard model presented in \S\ref{subsec:hierarchy}. 

 However, simple technicolor does not explain the origin of quark and lepton masses, 
because no Yukawa couplings are generated between {Higgs fields} and 
quarks or leptons.  Consequently, technicolor serves as a reminder 
that \textit{particle physics confronts two problems of mass:} explaining the masses of the 
gauge bosons, which demands an understanding of electroweak symmetry 
breaking; and accounting for the quark and lepton masses, which 
requires not only an understanding of electroweak symmetry breaking 
but also a theory of the Yukawa couplings that set the scale of 
fermion masses in the standard model.

To endow the quarks and leptons with mass, it is necessary to embed technicolor in a larger \textit{extended technicolor}
framework~\cite{Eichten:1979ah,Dimopoulos:1979es,Farhi:1980xs}  containing degrees of freedom that communicate the broken electroweak symmetry to the (technicolor-singlet) standard-model fermions. Specific implementations of these ideas face phenomenological challenges pertaining to flavor-changing neutral currents, the large top-quark mass, and precision electroweak measurements, but the idea of dynamical symmetry breaking remains an important alternative to the standard elementary scalar. For reviews and a summary of recent developments, see~\cite{Lane:2002wv,Hill:2002ap,Foadi:2007ue,Shrock:2007km}. 

Other suggestive work in the area of dynamical symmetry breaking also builds on the metaphor of the BCS theory of superconductivity, but attributes a special role to quarks of the third generation or beyond. 
A rich line, based on the notion that a top-quark condensate drives electroweak symmetry breaking, was initiated in~\cite{NambuTop,Miransky:1989ds,Bardeen:1989ds}. The idea that condensation of a strongly coupled fourth generation of quarks could trigger electroweak symmetry breaking is a lively area of contemporary research~\cite{Burdman:2007sx,Burdman:2008qh}.

\subsubsection{Other mechanisms for electroweak symmetry breaking}
Very informative surveys of new approaches to electroweak symmetry breaking are given in~\cite{Grojean:2007zz,Burdman:2008jm}. 

Much model building has occurred around the proposition that  the Higgs boson is a pseudo-Nambu-Goldstone boson of a spontaneously broken approximate global symmetry, with the explicit breaking of this symmetry collective in nature, that is, more than one coupling at a time must be turned on for 
the symmetry to be broken. These ``Little Higgs'' theories feature weakly coupled new physics at the TeV scale~\cite{Schmaltz:2005ky,Perelstein:2005ka}. When supplemented with a new symmetry called $T$-parity, under which new heavy particles are odd and standard-model particles are even, the Little Higgs theories can survive precision electroweak constraints and proffer a dark-matter candidate~\cite{Cheng:2004yc}. 

New ways of thinking about electroweak symmetry breaking arise when we contemplate the possibility that spacetime has more than the canonical four dimensions. 
Among the possibilities are models without a physical Higgs scalar, in which electroweak symmetry is hidden by boundary conditions~\cite{Csaki:2003dt,Csaki:2003zu,Csaki:2005vy}. The unitarity violation (\cf \S\ref{subsec:tevscale}) that would other be present in a theory without a Higgs boson is softened---deferred to energy scales well above $1\tev$---by the exchange of Ka\l{}uza~\cite{Kaluza:1921tu}--Klein~\cite{Klein:1926tv} (KK) excitations~\cite{AppChoFre} of standard-model particles such as the $W$. In this case, the KK recurrences constitute the new physics on the 1-TeV scale required by the general argument.

Suppose instead that the electroweak gauge theory is itself formulated in more than four dimensions. From our four-dimensional perspective, components of the gauge fields along the supplemental directions will be seen as scalar fields with respect to the conventional four-dimensional coordinates~\cite{Quiros:2003gg,Sundrum:2005jf}.

It is even conceivable that the electroweak phase transition is an emergent phenomenon arising from strong dynamics among the weak gauge bosons~\cite{Chanowitz:2004gk}.  If we take
the mass of the Higgs boson to very large values (beyond $1\tev$ in the
Lagrangian of the electroweak theory), the scattering among gauge bosons
becomes strong, in the sense that $\pi\pi$ scattering becomes strong on the
GeV scale, as we saw in \S\ref{subsec:tevscale}.  In that event, it is
reasonable to speculate that resonances form among pairs of gauge bosons,
multiple production of gauge bosons becomes commonplace, and that resonant
behavior could hold the key to understanding what hides the electroweak
symmetry.

\section{INCOMPLETENESS OF THE ELECTROWEAK THEORY \label{sec:SMincomplete}}
For all its successes, the electroweak theory leaves many questions unanswered. It does not explain the negative coefficient $\mu^2 < 0$ of the quadratic term in (\ref{eq:SSBpot}) required to hide the electroweak symmetry, and it merely accommodates, but does not predict, fermion masses and mixings. The Cabibbo--Kobayashi--Maskawa framework describes what we know of \textsf{CP} violation, but does not explain its origin. 
The discovery of neutrino flavor mixing, with its implication that neutrinos have mass, calls for an extension of the electroweak theory set out in  \S\ref{sec:ewtheory}. Moreover, an elementary Higgs sector is unstable against large radiative corrections. A pervasive nonzero vacuum expectation value for the Higgs field implies a uniform energy density of the vacuum that seems incompatible with observations. Neutrinos are the only dark-matter candidates within the standard model. They appear to contribute only a small share of the inferred dark-matter energy density, and as relativistic (``hot'') dark matter, not the cold dark matter required for structure formation in the early Universe. The \textsf{CP} violation observed in the quark sector, in accord with the CKM paradigm, seems far too small to account for the excess of matter over antimatter in the Universe.

\subsection{The problem of identity \label{subsec:ID}}
The structure of the \smgg\ standard model is of quantum chromodynamics and the electroweak theory, as we recalled it in \S\ref{sec:ewtheory}, can be written down in a few lines. But to calculate physical processes within the standard model---to apply the standard model to the real world---we need to specify at least 26 parameters. As tokens for the coupling parameters of the three factors of the gauge group, we may choose the strong coupling constant $\alpha_s$, the fine structure constant $\alpha_{\mathrm{em}}$, and the weak mixing parameter $\sin^2\theta_{\mathrm{W}}$. Two parameters are required to specify the shape of the Higgs potential \eqn{eq:SSBpot}. Then there are six quark masses and four parameters (three mixing angles and the \textsf{CP}-violating phase) of the CKM matrix \eqn{eq:ckmmatrix}. The charged leptons and (massive) neutrinos add six more mass parameters, three more mixing angles, and one more  \textsf{CP}-violating phase. Adding the (QCD) vacuum phase implicated in the strong \textsf{CP} problem brings the total to 26. Two more \textsf{CP}-violating phases enter if the neutrinos are their own antiparticles---Majorana particles.
At least 20 of these parameters are related to the physics of flavor.

The operational question, ``What determines the masses and mixings of the quarks and leptons?'' can be restated more evocatively, ``What makes a top quark a top quark, an electron an 
electron, and a neutrino a neutrino?'' 
 It is not enough 
to answer, ``The Higgs mechanism,'' because the fermion masses are 
a very enigmatic element of the electroweak theory.
 Once the electroweak 
symmetry is hidden, the electroweak theory accommodates fermion masses, 
but the values of the masses are set by the apparently 
arbitary couplings of the Higgs boson to the fermions (\cf Figure~\ref{fig:yuks}). Nothing 
in the electroweak theory is ever going to prescribe those couplings. 
It is not that the calculation is technically challenging; \textit{there is 
no calculation.}
Neutrino masses can be generated through Yukawa couplings, and in new ways as
well, because the neutrino may be its own antiparticle~\cite{Mohapatra:2004ht}.  

\begin{figure}[tb]
\begin{center}
\centerline{\includegraphics[width=0.5\columnwidth]{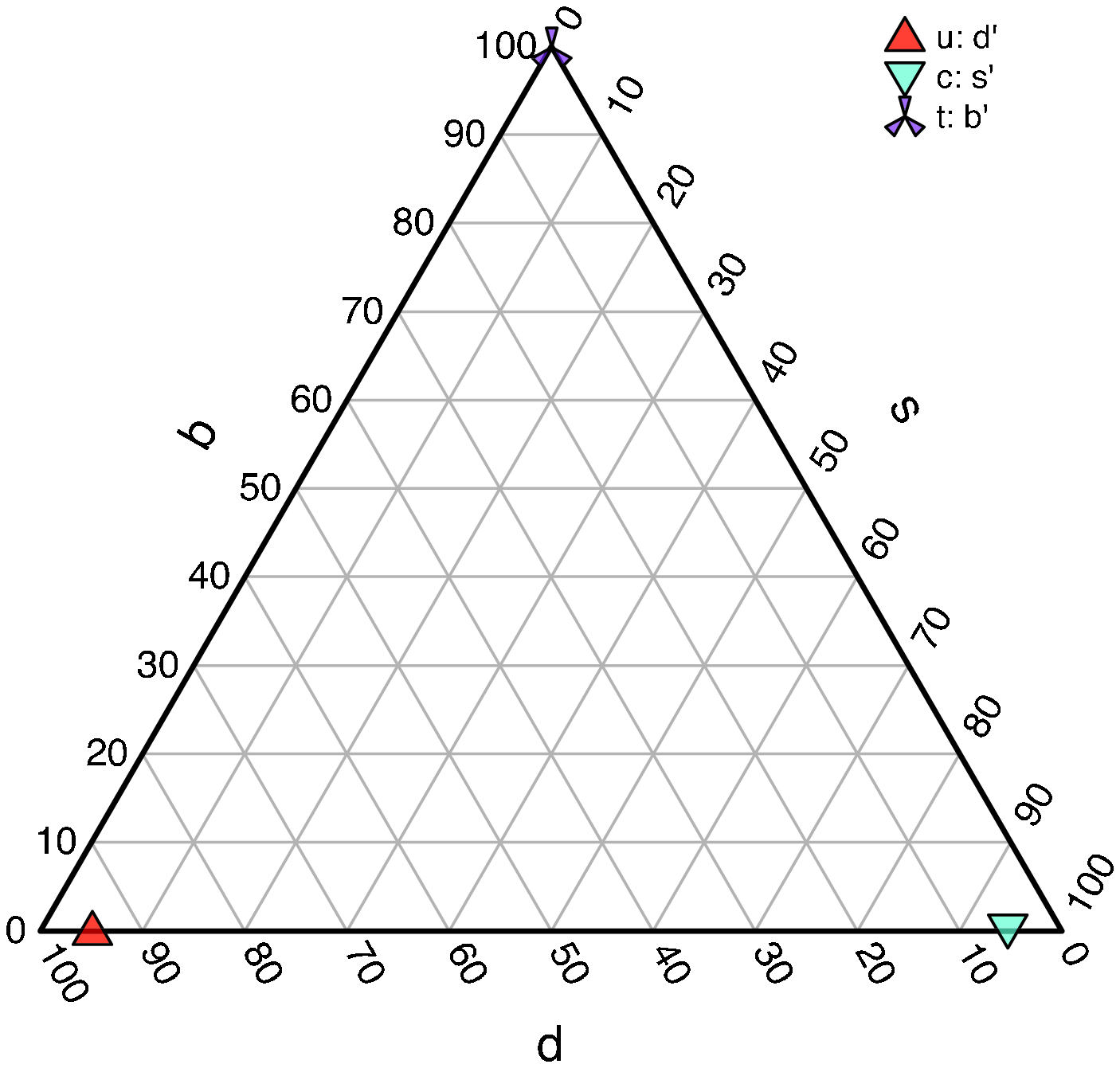} \quad\includegraphics[width=0.5\columnwidth]{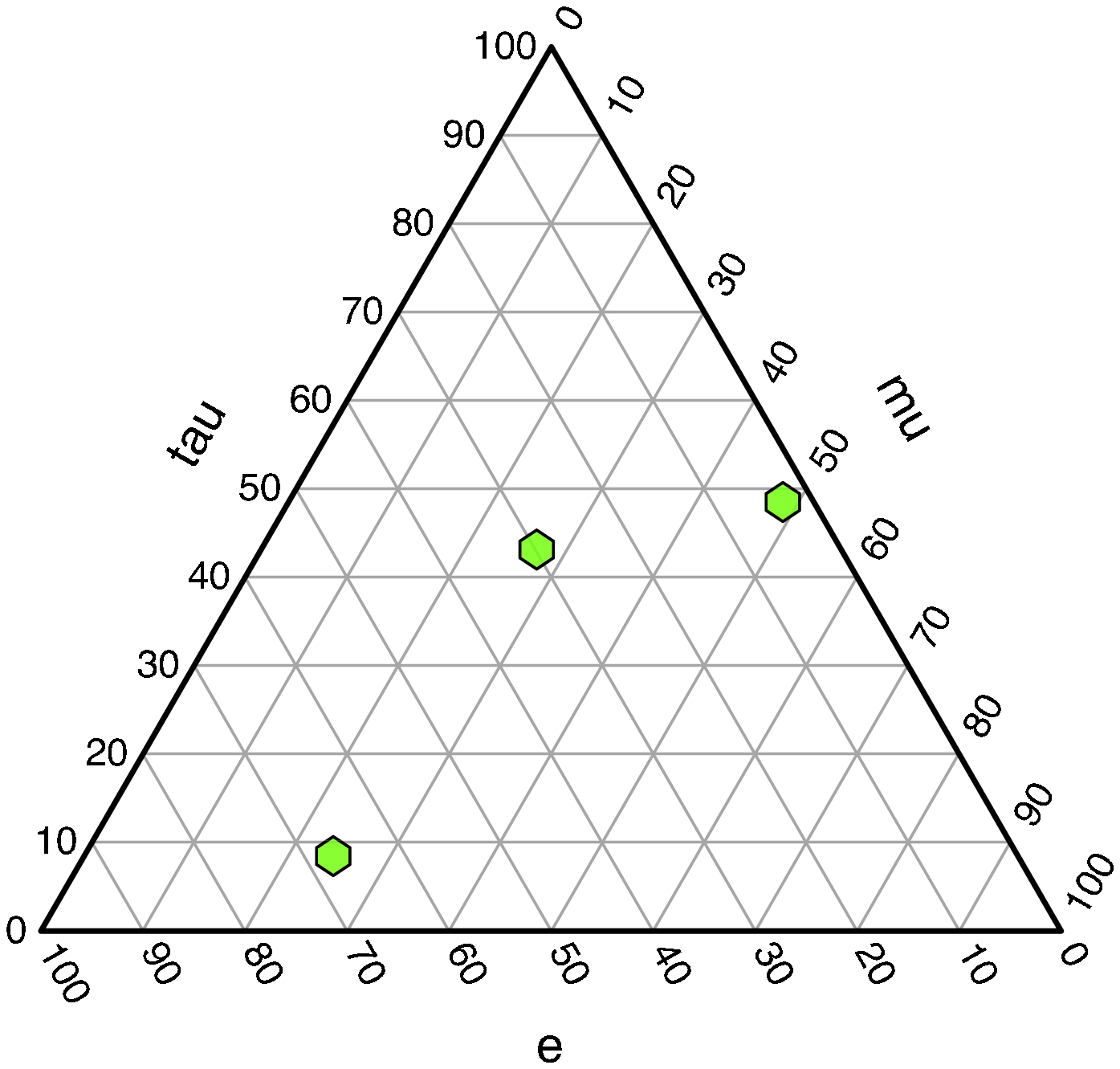}}
\caption{Left pane: $d, s, b$ composition of the quark flavor eigenstates $d^\prime$ (red $\bigtriangleup$), $s^\prime$ (green $\bigtriangledown$), $b^\prime$ (violet tripod).
Right pane: $\nu_e, \nu_\mu, \nu_\tau$ flavor content of the neutrino mass eigenstates $\nu_1, \nu_2, \nu_3$. The green hexagons denote central values, neglecting \textit{CP} violation in the lepton sector, and with the ``small'' mixing angle taken as  $\theta_{13} = 10^\circ$~\cite{Quigg:2008ki}. \label{fig:flavormixing}}
\end{center}
\end{figure}

Within the standard electroweak theory, it is not only the fermion masses, but also the mixing angles that parametrize the mismatch between flavor eigenstates and mass eigenstates, that are set by the Yukawa couplings.  
The  family relationships captured in the (Cabibbo--Kobayashi--Maskawa) quark mixing matrix are displayed in the ternary plot in the left pane of Figure~\ref{fig:flavormixing}. 
The coordinates are given by the squares of the CKM matrix elements in each row of \eqn{eq:ckmmags}. The $u$-quark couples mostly to $d$, $c$ mostly to $s$, and $t$ almost exclusively to $b$.

Our current knowledge of neutrino oscillations suggests the flavor content of the neutrino mass eigenstates depicted in the right pane of Figure~\ref{fig:flavormixing}. The pattern is very different from that of the quark sector: the mass eigenstate  $\nu_3$ consists of nearly equal parts of $\nu_{\mu}$ and $\nu_{\tau}$, perhaps with a trace of $\nu_e$, while $\nu_2$ contains similar amounts of $\nu_e$, $\nu_{\mu}$, and $\nu_{\tau}$, and $\nu_1$ is rich in $\nu_e$, with approximately equal minority parts of $\nu_{\mu}$ and $\nu_{\tau}$. Here $\nu_1$ is the lighter of the solar pair, $\nu_2$ is its heavier solar partner, and $\nu_3$ lies either above (normal hierarchy) or below (inverted hierarchy) the solar pair in mass.

The exciting prospect is that quark and lepton masses, mixing 
angles, and subtle differences in the behavior of particles and their antiparticles put us in contact with 
physics beyond the standard model. One important step toward understanding will be to ascertain whether the Higgs boson is indeed the agent behind fermion mass. Another will be to determine whether the light neutrinos are in fact their own antiparticles, as would be signaled by the observation of neutrinoless double-$\beta$ decay. Perhaps we find it hard to decode the message in the fermion masses and mixings because we are only seeing part of a larger picture, and that it will take discovering the spectrum of a new kind of matter---a fourth generation, or superpartners, or something entirely different---before it all begins to make sense.  

\subsection{The problem of widely separated scales \label{subsec:hierarchy}}
\subsubsection{The hierarchy problem \label{subsubsec:hier}}
Beyond the classical approximation, scalar mass parameters receive 
quantum corrections from loops that contain particles of spins 
$J=0, \cfrac{1}{2}$, and $1$, symbolically
\begin{displaymath}
M_H^2(p^2) = M_H^2(\Lambda^2) +\raisebox{-2.8ex}{\includegraphics[width=0.55\columnwidth]{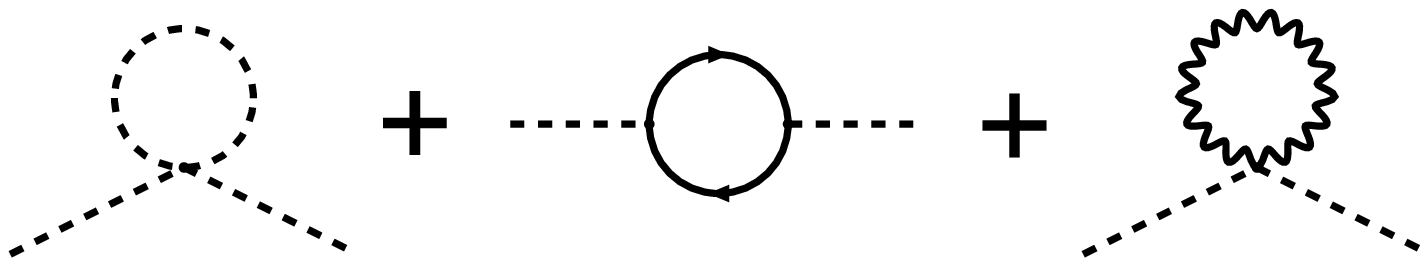}},
\label{eq:mechancete}
\end{displaymath}
where $\Lambda$ defines a reference scale at which the value of 
$M_H^2$ is known. 
The dashed lines represent the Higgs boson, solid lines with arrows represent fermions and antifermions, and wavy lines stand for gauge bosons.
The quantum corrections that determine the running mass lead potentially to divergences, 
\begin{equation}
	M_H^2(p^2) = M_H^2(\Lambda^2) + \mathcal{C}g^2\int^{\Lambda^2}_{p^2}dk^2 
	+ \cdots \;,
	\label{longint}
\end{equation}
where $g$ is the coupling constant of the theory, and the 
coefficient $\mathcal{C}$ is calculable in any particular theory.  
The loop integrals appear to be quadratically divergent, $\propto \Lambda^2$. 
In the absence of new physics, the reference scale $\Lambda$ would naturally be large.
If the fundamental interactions are described by quantum
chromodynamics and the electroweak theory, then a 
natural reference scale is the Planck mass,
$\Lambda \sim M_{\rm Planck}  = 
	\left({\hbar c}/{G_{\mathrm{Newton}}}\right)^{1/2} \approx 1.2 
	\times 10^{19}\gev$.
In a unified theory of the strong, weak, and electromagnetic 
interactions, a natural scale is the unification scale,
$\Lambda \sim U \approx 10^{15}\hbox{ - }10^{16}\gev$.
 Both estimates are very large compared to the electroweak scale, and so imply a very long range of integration. 
 
In order for the mass shifts 
induced by quantum corrections to remain modest, either something must limit the range of integration, or new physics must damp the integrand.
 The challenge of preserving widely separated electroweak and reference scales in the presence of quantum corrections is known as the \textit{hierarchy problem.} Unless we suppose that $M_H^2(\Lambda^2)$ and the quantum corrections are finely tuned to yield $M_H^2(p^2) \lesssim (1\tev)^2$, some new physics---a new symmetry or new dynamics---must intervene at an energy of approximately $1\tev$ to tame the integral in \Eqn{longint}. 
 
 Let us review the argument for the hierarchy problem: The unitarity argument (\cf \S\ref{subsec:tevscale}) showed that new physics must be present on the \onetev, either in the form of a Higgs boson, or other new phenomena. But a low-mass Higgs boson is imperiled by quantum corrections. New physics not far above the \onetev\ could bring the reference scale $\Lambda$ low enough to mitigate the threat. That is what happens in models of large~\cite{ArkaniHamed:1998rs,Antoniadis:1998ig} or warped~\cite{Randall:1999ee,Randall:1999vf} extra dimensions~\cite{Csaki:2004ay}, in which $M_{\rm Planck}$ is seen as a mirage, based on a mistaken extrapolation of Newton's law of gravitation to very short distances, or a new cutoff emerges, set by the scale of the extra dimension.

 If the reference scale is indeed very large, then either various contributions to the Higgs-boson mass must be precariously balanced or new physics must control the contribution of the integral in \Eqn{longint}. It is important to keep in mind that fine-tuning, perhaps guided by environmental selection, might be the way of the world~\cite{WeinbergMulti}. However, experience teaches us to be alert for symmetries or dynamics behind precise cancellations.

A new symmetry, not present in the standard model, could resolve the hierarchy problem.
  Exploiting the fact 
that fermion loops contribute with an overall minus sign relative to boson loops (because of 
Fermi statistics), \textit{supersymmetry}~\cite{Martin:1997ns,Djouadi:2005gj} balances the contributions of fermion and boson loops.  In unbroken supersymmetry, the 
masses of bosons are degenerate with those of their fermion 
counterparts, so the cancellation is exact. If supersymmetry is present in our world, it must be broken.
The contribution of the integrals may still be acceptably small if the 
fermion-boson mass splittings $\Delta M$ are not too large.  The 
condition that $g^2\Delta M^2$ be ``small enough'' leads to the 
requirement that superpartner masses be less than about 
$1\tev$. It is provocative to note that, with superpartners at $\mathcal{O}(1\tev)$, the $\mathrm{SU(3)_c}\otimes \ewgg$ coupling constants run to a common value at a unification scale of about $10^{16}\gev$~ \cite{Amaldi:1991cn}.

Theories of dynamical symmetry breaking (\cf \S\ref{subsubsec:techni}) offer a  second solution to the problem of the enormous range of integration in 
\eqn{longint}. In technicolor models, the Higgs boson is composite, and its internal structure comes into play  on the scale of its binding, $\Lambda_{\mathrm{TC}} \simeq 
\mathcal{O}(1~{\rm TeV})$. The integrand is damped, the effective range of integration is cut off, and 
mass shifts are under control.

A recurring hope among theorists has been the notion that the Higgs boson might be naturally light because it is the pseudo-Nambu--Goldstone boson (pNGB) of some approximate global symmetry. 
``Little Higgs" models~\cite{Schmaltz:2005ky,Perelstein:2005ka,Cheng:2004yc} introduce additional gauge bosons, vector-like quarks, and scalars on the TeV scale. These conspire, thanks to a global symmetry, to cancel the quadratic divergences in \eqn{longint} that result from loops of standard-model particles and defer the hierarchy problem to about $10\tev$. In contrast to supersymmetry, the cancellations arise from loops containing particles of the same spin. In ``twin Higgs'' models~\cite{Chacko:2005pe}, the new states do not carry standard-model charges.  The new physics at $\sim 10\tev$ raises impediments to conventional hopes for perturbative unification of the strong, weak, and electromagnetic interactions. Gauge-Higgs unification models based on warped five-dimensional geometry incorporate the pNGB interpretation of the Higgs boson and can exhibit logarithmic running of the gauge couplings that would support perturbative unification, as explained carefully in~\cite{Gherghetta:2006ha}.

\subsubsection{Tension between global fits and no new phenomena \label{subsubsec:tension}}
A fine-tuning problem may be seen to arise even when the scale $\Lambda$ is not extremely large. What has been called the ``LEP Paradox''~\cite{Barbieri:2000gf,Burdman:2006tz} refers to a tension within the precise measurements of electroweak observables carried out at LEP and elsewhere. On the one hand, the global fits summarized in Figure~\ref{fig:blueband} point to a light standard-model Higgs boson. On the other hand, a straightforward effective-operator analysis of possible beyond-the-standard-model contributions to the same observables gives no hint of any new physics---of the kind needed to resolve the hierarchy problem---below about $5\tev$.

 \begin{figure}[tb]
\begin{center}
\includegraphics[width=\columnwidth]{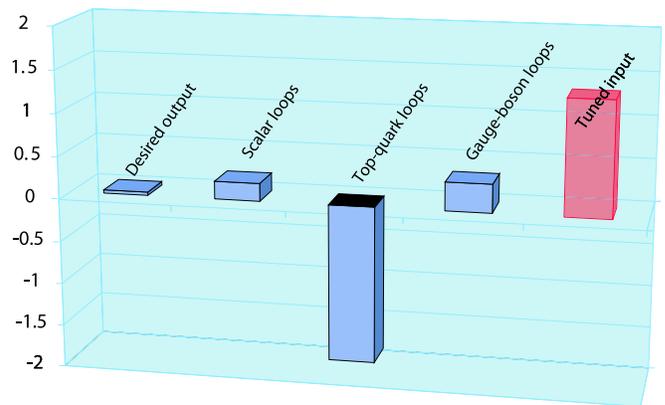}
\caption{Relative contributions to $\Delta M_H^2$ for a modest value of the cutoff parameter, $\Lambda = 5\tev$, in (\ref{longint}). \label{fig:finetune2}}
\end{center}
\end{figure}
Figure~\ref{fig:finetune2} shows that even with a cutoff $\Lambda = 5\tev$, a careful balancing act is required to maintain a small Higgs-boson mass in the face of quantum corrections, within the standard model, for which
\begin{equation}
\delta M_H^2 = {\displaystyle \frac{G_{\mathrm{F}} \Lambda^2}{4\pi^2\sqrt{2}}}(6M_W^2 + 3M_Z^2 + M_H^2 -12m_t^2) .
\label{eq:hiereq}
\end{equation}
 The chief cause for concern is the large contribution from the top-quark loop, 
 \begin{equation}
\left.\delta M_H^2\right|_{t\mathrm{-loop}} \approx  -\frac{3G_{\mathrm{F}}}{\pi^2\sqrt{2}}m_t^2\Lambda^2  \approx - 0.075\,\Lambda^2 .
\label{eq:toploop}
\end{equation}
We are left to ask what enforces the balance, or how we might be misreading the evidence.

\subsection{The vacuum energy problem \label{subsec:vacen}}
The cosmological constant problem---why empty space is so nearly massless---is one of the great mysteries of science~\cite{Weinberg:1988cp,Frieman:2008sn}. It is the reason why gravity has weighed on the minds of electroweak theorists, despite the utterly negligible role that gravity plays in particle reactions. Recall that the gravitational attraction between an electron and proton is forty-one orders of magnitude smaller than the electrostatic attraction at the same separation.

At the vacuum expectation value $\vev{\phi}$ of the Higgs field, the (position-independent) value of the Higgs potential is 
\begin{equation}
    V(\vev{\phi^{\dagger}\phi}) = \frac{\mu^{2}v^{2}}{4} = 
    - \frac{\abs{\lambda}v^{4}}{4} < 0.
    \label{minpot}
\end{equation}
Identifying $M_{H}^{2} = -2\mu^{2}$, we see that the Higgs potential 
contributes a uniform vacuum energy density,
\begin{equation}
    \varrho_{H} \equiv \frac{M_{H}^{2}v^{2}}{8}.
    \label{eq:rhoH}
\end{equation}
From the perspective of general relativity, this amounts to adding a cosmological constant, $\Lambda = (8\pi G_{\mathrm{N}}/c^{4})\varrho_{H}$, to Einstein's equation, where $G_{\mathrm{N}}$ is Newton's gravitational constant~\cite{Linde:1974at,Veltman:1974au,Peebles:2002gy}.  

Recent observations of the accelerating expansion of the Universe~\cite{Riess:1998cb,Perlmutter:1998np} raise the intriguing possibility that the cosmological constant may be different from zero, but the essential fact is that the observed vacuum energy density must be very small indeed~\cite{Komatsu:2008hk},
\begin{equation}
    \varrho_{\mathrm{vac}} \lesssim 10^{-46}\gev^{4} \approx (\hbox{a few meV})^4\; .
    \label{eq:rhovaclim}
\end{equation}
Therein lies the puzzle: if we take
$v = (G_{\mathrm{F}}\sqrt{2})^{-\frac{1}{2}}  \approx 246\gev$  
and insert the current experimental lower bound~\cite{Barate:2003sz} 
$M_{H} \gtrsim 114.4\gev$ into \eqn{eq:rhoH}, we find that the Higgs field's
contribution to the vacuum energy density is
\begin{equation}
    \varrho_{H} \gtrsim  10^{8}\gev^{4},
    \label{eq:rhoHval}
\end{equation}
some 54 orders of magnitude larger than the upper bound inferred from 
the cosmological constant. This mismatch has been a source of dull headaches for more than three decades.

The problem is still more serious in a unified theory of the strong, 
weak, and electromagnetic interactions, in which other (heavy!) Higgs fields 
have nonzero vacuum expectation values that may give rise to still 
larger vacuum energies.  At a fundamental level, we can therefore conclude 
that a spontaneously broken gauge theory of the strong, weak, and 
electromagnetic interactions---or merely of the electroweak 
interactions---cannot be complete. The vacuum energy problem must be an important 
clue.  But to what?

The tentative evidence for  a nonzero cosmological constant recasts the problem in two important ways.
First, instead of looking for a principle that would forbid a cosmological constant, perhaps a symmetry principle that would set it exactly to zero, we may be called upon to explain a tiny cosmological constant.
Second, if the interpretation of the accelerating expansion in terms of dark energy is correct, we now have observational access to some new stuff whose equation of state and other properties we can try to measure.  Maybe that will give us the clues that we need to solve this old problem, and to understand how it relates to the electroweak theory.

\subsection{Lacunae \label{subsec:missing}}
The electroweak theory is unresponsive to some questions that are inspired by observations of the Universe at large. 

\subsubsection{Dark matter \label{subsubsec:dm}}
The rotation curves of spiral galaxies and supporting evidence from the cosmic microwave background and large-scale structure point to ``dark matter'' that makes up 25\% of the Universe's energy density~\cite{Komatsu:2008hk}.  An appealing interpretation is that the dark matter is composed of one or more neutral relics from the early Universe. Within the standard model, the only candidates are neutrinos, for which the weight of experimental and observational evidences argues for masses smaller than about $1\ev$.

Using the calculated number density of $56\cm^{-3}$ for each $\nu$ and $\bar{\nu}$ flavor in the current universe, we can deduce the neutrino contribution to the mass density, expressed in units of the critical density, as $\rho_{\mathrm{c}} \equiv 3H_0^2/8\pi G_{\mathrm{N}} = 1.05 h^2 \times 10^4\ev\cm^{-3} = 5.6 \times 10^3\ev\cm^{-3}$, where $H_0$ is the Hubble parameter now, $G_{\mathrm{N}}$ is Newton's constant, and I have taken the reduced Hubble constant to be $h = 0.73$~\cite{Quigg:2008ab}. Neutrinos contribute a normalized energy density $\Omega_\nu \gtrsim (1.2, 2.2) \times 10^{-3}$ for the (normal, inverted) spectrum, and no more than 10\% of critical density, should the lightest neutrino mass approach $1\ev$. Neutrinos are not, however, candidates for the cold dark matter (nonrelativistic at the time of structure formation) that is favored by scenarios for structure formation in the Universe~\cite{Ratra:2007sa,DarkMatter,Hooper:2009zm}.

\subsubsection{Baryon asymmetry of the Universe \label{subsubsec:bau}}
Why does matter dominate over antimatter in the observable Universe~\cite{QuinnNir}? Observations indicate that the density of antibaryons is negligible, whereas the average density of baryonic matter, characterized by the baryon-to-photon ratio
\begin{equation}
\eta \equiv \frac{n_B - n_{\bar{B}}}{n_\gamma} = (6.14 \pm 0.25) \times 10^{-10} .
\label{eq:Basym}
\end{equation}
where $n_B$, $n_{\bar{B}}$, and $n_\gamma$ are respectively the baryon, antibaryon, and photon number densities. Cosmological observations, anchored by the WMAP measurements of the Doppler peaks of temperature fluctuations in the cosmic microwave background radiation~\cite{Komatsu:2008hk}, imply that the current normalized baryon energy density is
\begin{equation}
\Omega_B = 0.0456 \pm 0.0015 ,
\label{eq:omegab}
\end{equation}
from which one can infer
\begin{equation}
\eta = (6.22 \pm 0.19) \times 10^{-10} .
\label{eq:etanow}
\end{equation}
Why is the ratio not zero? In an inflationary cosmology, conventional processes should produce equal numbers of baryons and antibaryons.

Three conditions are required to generate a baryon asymmetry out of neutral initial conditions~\cite{Sakharov:1967dj}: (i) the existence of fundamental processes that violate baryon number; (ii) microscopic \textsf{CP} violation; and (iii) departure from thermal equilibrium during the epoch in which baryon-number violating processes were important. A clear and compact survey of our current understanding of the baryon number of the Universe appears in~\cite{Cline:2006ts}. How well does the electroweak theory respond? 

The nonequilibrium condition is met by the expanding Universe. 
The electroweak theory does contain \textsf{CP} violation, in the CKM framework.
At the level of perturbation theory, the electroweak theory conserves baryon number $B$ and lepton number $L$, but that is not the case in the nonperturbative realm.
Weak \wigg\  instantons  violate $B$ and $L$,
conserving $B-L$, but have a negligibly small effect at temperatures  $T$ much lower
than the electroweak scale $v \approx 246\gev$. Their contributions to physical
processes are suppressed by the factor $\exp(-8 \pi^2/g^2)$ at zero temperature, where $g$ is the \wigg\ gauge coupling~\cite{'tHooft:1976up}.  For $T \gtrsim v$, the sphalerons tend to erase a pre-existing baryon asymmetry of the Universe, and could under some conditions generate a significant baryon asymmetry. Our best assessment is that electroweak baryogenesis, within the standard model, well falls short of explaining the ratio \eqn{eq:etanow}. Some new physics, beyond the standard model, is required. A popular hypothesis is \textit{leptogenesis,} in which the baryon asymmetry of the Universe is produced from a lepton asymmetry generated in the decays of a heavy sterile neutrino~\cite{Buchmuller:2005eh,Davidson:2008bu}.

\subsubsection{Quantization of electric charge \label{subsubsec:Qquant}}
The proton and electron charges balance to an astonishing degree~\cite{Amsler:2008zzb}:
\begin{equation}
\abs{Q_p + Q_e} < 10^{-21}\abs{Q_e}.
\label{eq:neutral}
\end{equation}
If there were no connection between quarks and leptons, since quarks make
up the proton, then the balance of the proton and electron charge would
just be a remarkable coincidence, which seems an unsatisfying explanation.  Some
principle must relate the charges of the quarks and the leptons.  What
is it?  An appealing strategy is to assign quarks and leptons to extended families. This is the approach of unified theories of the strong, weak, and electromagnetic interactions. It carries the implication of interactions that transform quarks into leptons, which has consequences for proton decay and baryogenesis. The idea that quantum chromodynamics and the electroweak theory might be unified is made plausible by the fact that both are gauge theories, with similar mathematical structure.

Another encouragement to consider quark-lepton unification comes from the electroweak theory itself. A world governed by \ewgg\ interactions and composed only of quarks, or only of leptons, would be anomalous~\cite{Bouchiat:1972iq}, in the technical sense that quantum corrections would break the gauge symmetry on which the theory is based.  In our left-handed world, an anomaly-free electroweak theory is possible only if weak-isospin pairs of color-triplet quarks accompany weak-isospin
pairs of color-singlet leptons.  For these reasons, it is nearly
irresistible to consider a unified theory that puts quarks and leptons
into a single extended family.

\subsubsection{Absence of gravity \label{subsubsec:nograv}}
The gravitational force is famously negligible in the realm of particle physics.  In the language of Feynman rules, dimensional analysis shows that the emission of a graviton is suppressed by a factor
\begin{equation}
E^\star/M_{\mathrm{Planck}} ,
\label{eq:gravem}
\end{equation}
where $E^\star$ is a characteristic energy of the transition.
  The Planck mass ($M_{\mathrm{Planck}} \equiv (\hbar  c/G_{\mathrm{Newton}})^{1/2}\approx 1.22
\times 10^{19}\gev$) is a big number because Newton's constant is small in the
appropriate units.  Except in special circumstances, such as the excitation of many extra-dimensional modes~\cite{Hewett:2002hv}, graviton emission need not be taken into account in particle physics.

However, we have seen in our discussion of the vacuum energy problem (\cf \S\ref{subsec:vacen}) that the relationship of the electroweak theory to gravitation cannot be ignored. The hierarchy problem (\cf \S\ref{subsec:hierarchy}) is acute, in part because the Planck scale is so distant from the electroweak scale. These connections, as well as the desire to understand why gravity is so much weaker than the \smgg\ interactions, motivate efforts to integrate gravity with the standard model. When imagining how this might be achieved, it is important to bear in mind that we have probed the electroweak
theory and QCD up to about $1\tev$, but we
have  tested the inverse-square law of gravity only down to distances  just shorter than $0.1\mm$, corresponding to energies of 10 \textit{milli}-electron volts!

\section{THE NEW ERA \label{sec:dawn}}
The Tevatron is expected to operate through 2011, producing a total of $10\fb^{-1}$ for analysis by the CDF and D0 collaborations. The experimenters are optimistic that a sample of that size will be sufficient---in the absence of a signal---to set a 95\% exclusion limit for the standard-model Higgs boson over the entire range currently favored by the global fits discussed in \S\ref{subsec:excon}. Barring a breakthrough in analysis techniques, discovery of the standard-model Higgs boson at 5-$\sigma$ significance is extremely unlikely at the Tevatron, unless the production rate should be enhanced (for example, by a fourth generation of quarks). At the interesting level of 3-$\sigma$ evidence, the situation is more promising. The experiments have quoted the odds of establishing ``evidence'' at about one in three for $120\gev \lesssim M_H \lesssim 145\gev$, and better than one in two for $M_H \lesssim 116\gev$ and $150\gev \lesssim M_H \lesssim 177\gev$~\cite{Dmitri}. At a minimum, we will know more about where the (standard-model) Higgs boson is not by the time the LHC Higgs search begins in earnest.

The parameters of the Large Hadron Collider are shaped by the imperative to make a thorough exploration of the 1-TeV scale, and so to elucidate the mechanism of electroweak symmetry breaking. But the LHC is a discovery machine, broadly understood, not limited to the search for the agent of electroweak symmetry breaking. See~\cite{Eichten:1984eu} for a general survey, and the ATLAS~\cite{Aad:2009wy} and CMS~\cite{Ball:2007zza} physics reports for the detector capabilities, with many specific illustrations.

\subsection{Electroweak questions for LHC experiments \label{subsec:ewqs}}
Will the new physics that we anticipate on the 1-TeV scale be a Higgs boson, in some guise, or new strong dynamics? If a Higgs boson, will there be one, or several, and will it---or they---turn out to be elementary or composite? Is the Higgs boson indeed light, as anticipated by the global fits to electroweak precision measurements? Does the Higgs boson give mass only to the electroweak gauge bosons, or does it also endow the fermions with mass? Proceeding step by step, does the ``$H$'' couple to fermions (a large $t\bar{t}H$ coupling might be inferred from its production rate)? Are the branching fractions for decays into fermion pairs in accord with the standard model? A difficult follow-up question, should we find that the Higgs boson \textit{is} responsible for fermion mass, is what determines the masses and mixings of the fermions?

The Higgs boson(s) could couple to particles beyond those known in the standard model. Does the pattern of Higgs-boson decays imply new physics? Will unexpected or rare decays of the Higgs boson reveal new kinds of matter? If more than one, apparently elementary, Higgs boson is found, will that be a sign for a supersymmetric generalization of the standard model, or for a different sort of two-Higgs-doublet model? What stabilizes the Higgs-boson mass below $1\tev$? How can a light Higgs boson coexist with the absence of signals for new phenomena? 
Is electroweak symmetry breaking an emergent phenomenon connected with strong dynamics? Is electroweak symmetry breaking related to gravity through extra spacetime dimensions?

If the new physics observed on the TeV scale is suggestive of new strong dynamics, how can we diagnose the nature of the new dynamics? What takes the place of a Higgs boson?

\subsection{More new physics on the TeV scale? \label{subsec:more}}
The partial-wave unitarity argument reviewed in \S\ref{subsec:tevscale} indicates that a thorough exploration of the TeV scale will produce important insights into the nature of electroweak symmetry breaking. At a strongly suggestive level, we have reason to expect additional new phenomena in this energy range.

The large gap between the electroweak scale and the unification scale or the Planck scale menaces the Higgs-boson mass with large quantum corrections that would lift it far above $1\tev$. If the required small mass of the Higgs boson is not stabilized by fine tuning, then new physics is needed on the TeV scale. Familiar examples are supersymmetry, with a spectrum of superpartners beginning to appear on the TeV scale, and technicolor, with its own spectrum of new technipions and $W^+W^-$ resonances. The common characteristic of solutions to the hierarchy problem is that they lead naturally to the expectation of new physics on the TeV scale. 
See~\cite{Barbieri:2008xg,Ellis:2009pz} for insightful surveys of some new-physics signatures to be anticipated at the LHC.

The evidence that cold dark matter is a significant component of the energy-density budget of the Universe is an independent argument for new phenomena on the TeV scale. As with the hierarchy problem, the implication is highly suggestive, but not inevitable: a few-hundred-GeV particle that couples with weak-interaction strength could supply the missing mass~\cite{Bertone:2004pz}.

To summarize, we have three indications for dramatic new developments on the TeV scale: the requirement for a Higgs boson or new dynamics, and the strong suggestions of new phenomena to solve the hierarchy problem and of particle dark matter. With the great discovery reach of the LHC, many other possibilities are open, including new heavy fermions and new force particles.
\subsection{How knowledge might accumulate}
How our understanding progresses in light of information from the LHC depends, of course, on what Nature has in store for us, and how our attention is attracted this way or that in light of discoveries. However, the extensive studies carried out in preparation for the ATLAS and CMS experiments, and informed by the Tevatron experience, allow us to anticipate the sensitivity required for various potential observations and discoveries. It is implicit that understanding of the detectors progresses as data are accumulated~\cite{Gianotti:2005fm,Froidevaux:2009dy}.

With an integrated luminosity of only $10\pb^{-1}$, the experiments will begin to characterize event structure in the new energy regime, to measure jet and di-jet spectra, and to study $J\!/\!\psi$, $\Upsilon$, and $W^\pm$ production. Within the first hundred days of stable running, at about $50\pb^{-1}$, $Z^0 \to \ell^+\ell^-$ comes into view, and $t\bar{t}$ pairs should be observed. This will be an important milestone, since top-quark events provide subtle training grounds for detector algorithms that will allow tops to be identified as physics objects, and represent a background that must be mastered for many new-physics searches. When the data set reaches approximately $100\pb^{-1}$, incisive measurements of standard-model parameters come into reach.

A few early discoveries are possible for data sets of a few hundred$\pb^{-1}$ at $\sqrt{s} = 14\tev$: a $Z^\prime$ representing a new force of Nature, with mass up to approximately $1\tev$, and light ($\approx 500\gev$) squarks and gluinos that would give evidence of supersymmetry. At about $1\fb^{-1}$, standard-model Higgs-boson physics opens up, first with discovery sensitivity for the channel $H \to ZZ \to \mu^+\mu^-\,\mu^+\mu^-$ with $M_H \approx 180\gev$. Establishing a Higgs-boson signal at low mass is more demanding: at $M_H \approx 115\gev$, an integrated luminosity $\gtrsim 5\fb^{-1}$ will be required. At that point, combined channels from the experiments will cover the full allowed range of standard-model Higgs-boson masses.

When the LHC data set passes roughly $10\fb^{-1}$, a spin-2 dilepton resonance characteristic of extra spacetime dimensions will be visible up to $1\tev$. An order of magnitude more gives a discovery reach for leptoquarks up to $1.5\gev$, and a sensitivity to a compositeness scale $\Lambda^* = 30\tev$. Integrated luminosity of $300\fb^{-1}$, representing approximately five years of LHC experimentation, should suffice for the observation of TeV-scale $W^+W^-$ resonances, and expand the discovery reach for squarks and gluinos up to $2.5\tev$.

\begin{figure}
\centerline{\includegraphics[width= \columnwidth]{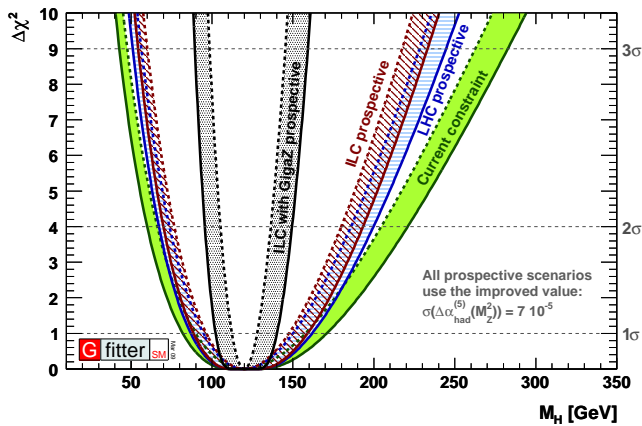}}
\caption{\textsf{Gfitter} constraints on the Higgs-boson mass obtained for four future scenarios. Parabolas in $\Delta\chi^2$ are shown with their theoretical error bands. From wider to narrower: present constraint, LHC expectation, ILC expectations excluding and including the Giga-$Z$ option~\cite{Flaecher:2008zq}.}
\label{fig:GfitterFuture}
\end{figure}

Understanding advances not only by discovery of new phenomena, but also by improvement of precision. As one example, Figure~\ref{fig:GfitterFuture} projects improvements in the global-fit constraints on the mass of the standard-model Higgs boson in light of measurements to be carried out at the LHC, and measurements that might be made at the proposed International Linear Collider, a 500-GeV electron-positron collider. Increasing the sample of $Z$-bosons by two orders of magnitude with the so-called Giga-$Z$ option for a linear collider further sharpens the constraint.
[For surveys of the full ILC physics program, see~\cite{Murayama:1996ec,Dawson:2004xz}.] As was the case for the values of $m_t$ inferred before the discovery of top (\cf Figure~\ref{fig:toptimeseries}), comparison of the global-fit constraints with the observed Higgs-boson mass will be an incisive test of the standard model.

\subsection{The Intensity Frontier}
Historically, much of the motivation for the electroweak theory came from detailed measurements at low energies, and such experiments have led the validation of the CKM structure of the charged-current weak interaction and established the suppression of flavor-changing neutral currents. The main imperative now is to explore the TeV scale, to establish the mechanism for electroweak symmetry breaking. That task will soon pass from the Tevatron to the Large Hadron Collider. Many interesting questions remain in flavor physics. The importance of intensity-frontier experiments for reshaping our understanding of particle physics can be enhanced by their conversation with the LHC experiments that explore the TeV scale. 

New sensitivity can bring surprises. The FCNC examples we saw in \S\ref{subsec:fcnc} show that there are good opportunities for physics beyond the standard model to appear. Prospects for the study of $B$, $D$ and $K$ decays are reviewed in~\cite{Buchalla:2008jp}. The lepton sector is taking on greater interest following the discovery of neutrino mixing and the possibility that \textsf{CP} violation might be observable in neutrino interactions~\cite{Freedman:2004rt}. Moreover, new searches for charged-lepton flavor violation, and for \textsf{CP} violation in lepton dipole moments offer the possibility of dramatic discoveries~\cite{Raidal:2008jk}. Interpretations of new phenomena observed at the LHC will be tested and refined by looking for the virtual effects of the new particles in rare processes studied at low energies. The LHC experiments, including LHC$b$, have their own role in flavor physics~\cite{delAguila:2008iz}.

\section{SUMMARY \label{sec:sum}}
Over the past two decades, experiments have tested the gauge sector and the flavor sector of the electroweak theory extensively, so that we may now regard the electroweak theory as a law of nature---subject, of course, to revision in the light of new evidence. Much of the experimental evidence was recounted in \S\ref{sec:law}. The body of evidence is both broad and deep, and while it leaves room for physics beyond the standard model, it also constrains new physics in significant ways.

Experiments at the Large Hadron Collider will probe the electroweak symmetry breaking sector on the 1-TeV scale, where we may also hope to find pointers to physics beyond the standard model.
The clues in hand suggest that the agent of electroweak symmetry breaking represents a novel fundamental interaction operating on the Fermi scale. \textit{We do not know what that force is.}

A leading possibility is that the agent of electroweak symmetry breaking is an elementary scalar, the Higgs boson of the electroweak standard model. Global fits to electroweak measurements indicate that the standard-model Higgs boson should be found with a mass not much more than $200\gev$.
An essential step toward understanding the new force that shapes our world is, therefore, to search for the Higgs boson and to explore its properties.

We have seen in \S\ref{subsec:Hint} that different sensitive observables prefer different values for the Higgs-boson mass. We do not know whether that reflects an unexceptional scatter or is a harbinger of physics beyond the standard model. In any case, it is important to search for the Higgs boson over the complete range of \textit{a priori} acceptable values, and the ATLAS and CMS experiments will accomplish that.

As we detailed in \S\ref{sec:SMincomplete}, the standard model is incomplete. It shows how the masses of the quarks and leptons might arise, but does not predict their values. It does not even give a qualitative understanding of why the quark-mixing parameters are small and hierarchical, nor why the pattern of neutrino mixing should be so different. For a provocative essay on one path to a more comprehensive understanding, see~\cite{Schellekens:2008kg}.

The hierarchy problem seems to require a solution in terms of dynamics or a symmetry---although fine tuning is a logical possibility. The vacuum energy problem indicates that something essential is missing in our understanding. These are problems within the electroweak theory. Other shortcomings, including the absence of a dark-matter candidate, speak to the limited reach of the electroweak theory.

We can be confident that the origin of gauge-boson masses will be understood on the TeV scale.  
We do not know where we will decode the pattern of the Yukawa couplings that set the fermion masses. 
However, candidate solutions to the hierarchy problem entail new physics on the TeV scale, and the weakly-interacting-massive-particle (WIMP) solution to the dark-matter question suggests a mass in the few-hundred-GeV range. These hints suggest that, in addition to the electroweak-symmetry-breaking physics we confidently expect to see at the LHC, there is every likelihood of more new phenomena.

The electroweak theory is a remarkable achievement. It gives a deeper understanding of two of the fundamental forces of nature---electromagnetism and the charged-current weak interaction---and adds the neutral-current weak interaction to the mix. It accounts for a wide variety of experimental measurements, and has survived many tests as a quantum field theory. It meets the most important criteria for a good theory: we get more out than we put in, and it raises new and significant questions. 

We are on the cusp of a new level of understanding, with the nature of electroweak symmetry breaking virtually certain to be revealed on the 1-TeV scale. At the same time, the incompleteness of the electroweak theory argues that we have much more to learn. Part of the high anticipation that attends the coming of the LHC is that the experimental opportunities on the TeV scale involve distinct problems that could well be related, and that might all be related through the electroweak theory. We will soon know how robust the connections are.
  
\begin{acknowledgments}
Fermilab is operated by the Fermi Research Alliance under contract no.\  DE-AC02-07CH11359 with the
U.S.\ Department of Energy. I thank Hans K\"{u}hn and Uli Nierste for a stimulating environment in Karlsruhe and acknowledge with pleasure the generous support of the Alexander von Humboldt Foundation. I am grateful to Luis \'{A}lvarez-Gaum\'{e} and other members of the CERN Theory Group for their hospitality. I thank Gustavo Burdman and Paddy Fox for perceptive comments on the manuscript.
\end{acknowledgments}

\bibliographystyle{utphysrevA}

\bibliography{EW4AR}

\end{document}